\documentstyle[12pt]{article}
\newcommand{\beq}{\begin{equation}}
\newcommand{\eeq}{\end{equation}}

\def\half{{\textstyle{1\over2}}}
\def\quarter{{\textstyle{1\over4}}}

\def\p1half{{\textstyle{{{p+1}\over{2}}}}}

\def\23phalf{{\textstyle{{{23-p}\over{2}}}}}

\def\quarter{{\textstyle{1\over4}}}
\begin{document}
\thispagestyle{empty}
\begin{titlepage}

\bigskip
\hskip 3.7in{\vbox{\baselineskip12pt
%\hbox{hep-th/0105244}
}}

\bigskip\bigskip
\centerline{\large\bf Thermal Duality Transformations and the
Canonical Ensemble} \centerline{\large\bf The Deconfining Long
String Phase Transition}

\bigskip\bigskip
\bigskip\bigskip
\centerline{\bf Shyamoli Chaudhuri \footnote{Current Address: 1312
Oak Drive, Blacksburg, VA 24060. Email: shyamolic@yahoo.com} }
\centerline{214 North Allegheny St.} \centerline{Bellefonte, PA
16823}
\date{\today}

\bigskip
\begin{abstract}
\vskip 0.1in We give a first principles formulation of the
equilibrium statistical mechanics of strings in the canonical
ensemble, compatible with the Euclidean timelike T-duality
transformations that link the six supersymmetric string theories
in pairs: heterotic $E_8$$\times$$E_8$ and ${\rm Spin}(32)/{\rm
Z}_2$, type IIA and type IIB, or type IB and type I$^{\prime}$. We
demonstrate that each of the supersymmetric string ensembles
exhibits a $T^2$ growth in the free energy at high temperatures
far above the string scale, precisely as conjectured by Atick and
Witten in 1988, and shown to follow as a consequence of thermal
self-duality in the closed bosonic string ensemble by Polchinski
more recently. We verify that the low energy field theory limit of
our expression for the string free energy reproduces the expected
$T^{10}$ growth when the contribution from massive string modes is
suppressed. In every case, heterotic, type I, and type II, we can
definitively rule out the occurrence of an exponential divergence
in the one-loop string free energy above some critical
temperature: {\em previous claims of evidence for the breakdown of
the supersymmetric string canonical ensembles at a limiting
Hagedorn temperature are incorrect}. Finally, we identify a
macroscopic loop amplitude in the type I string theories which
yields the expectation value of a single Wilson-Polyakov-Susskind
loop in the low energy finite temperature supersymmetric gauge
theory limit, an order parameter for a thermal phase transition at
a string scale temperature. We point out that precise computations
can nevertheless be carried out on either side of the phase
boundary by using the low energy finite temperature supersymmetric
gauge theory limits of the pair of thermal dual string theories,
type IB and type I$^{\prime}$. Note Added (Sep 2005).
\end{abstract}

\end{titlepage}

\section{Introduction}

\vskip 0.1in T-duality invariance, the result of interchanging
small radius with large radius, $R$ $\to$ $\alpha'/R$, is a {\em
spontaneously} broken symmetry in String/M Theory: in other words,
a T-duality transformation on an embedding target space coordinate
will, in general, map a given background of String/M theory to a
different background of the same theory. To be specific, the
circle-compactified $E_8$$\times$$E_8$ heterotic oriented closed
string theory is mapped under a T-duality to the
circle-compactified ${\rm Spin}(32)/{\rm Z}_2$ heterotic oriented
closed string theory \cite{ky,nair,busch,gv,itoyama,ginine}. The
circle-compactified type IIA oriented closed string theory is
mapped to the circle-compactified type IIB oriented closed string
theory \cite{din1,dlp}. And, finally, the unoriented type IB open
and closed string theory is mapped under a T-duality
transformation to the, rather unusual, type I$^{\prime}$
unoriented open and closed string theory: the integer-moded open
string momentum modes associated with the compact coordinate are
mapped to the integer-moded closed string winding modes in the
T-dual type I$^{\prime}$ theory \cite{dlp,dbrane,polchinskibook}.

\vskip 0.1in In this paper we will examine the consequences of
such T-duality transformations on the Wick-rotated time coordinate
$X^{0}$. It is clear that a Wick rotation on $X^0$ maps the
noncompact $SO(9,1)$ Lorentz invariant background of a given
supersymmetric string theory to the corresponding $SO(10)$
invariant background, with an embedding time coordinate of
Euclidean signature. The Wick-rotated $SO(10)$ invariant
background arises naturally in any formulation of equilibrium
string statistical mechanics in the canonical ensemble, the
statistical ensemble characterized by fixed temperature and fixed
spatial volume $(\beta ,V)$. The Polyakov path integral over
connected world-surfaces is formulated in a target spacetime of
fixed spacetime volume. Thus, the one-loop vacuum functional in
the $SO(10)$ invariant background computes precisely the sum over
connected one-loop vacuum graphs in the target space $R^9$ in the
finite temperature vacuum at temperature $T$ \cite{poltorus}. The
appearance of a {\em tachyonic} mode in the string thermal
spectrum is an indication that the worldsheet conformal field
theory is no longer at a fixed point of the 2d Renormalization
Group (RG): the tachyon indicates a relevant flow of the 2d RG.
The question of significance is then as follows: does the relevant
flow terminate in a new infrared fixed point? If so, the new fixed
point determines the true thermal string vacuum. An equilibrium
statistical mechanics of strings requires that this fixed point
belong to a {\em fixed line} parameterized by inverse temperature
$\beta$ \cite{cs,relevant}: the precise analog under Wick rotation
of the line of fixed points parameterized by the radius of a
compact spatial coordinate in the $SO(9,1)$ vacuum.

\vskip 0.1in Target spacetime supersymmetry, and its spontaneous
breaking in the thermal vacuum along the line of fixed points
parameterized by $\beta$, introduces new features into this
discussion. We must require compatibility with the expected
properties of the low energy field theory limit where the
contribution from massive string modes has been suppressed,
namely, those of a 10D finite temperature supersymmetric gauge
theory. We must also require consistency with string theoretic
symmetries of both worldsheet, and target space, origin in the
Wick rotated $SO(10)$ invariant background. In particular,
Euclidean T-duality transformations must link the thermal vacua of
the six different supersymmetric string theories in pairs:
heterotic $E_8$$\times$$E_8$ and ${\rm Spin}(32)/{\rm Z}_2$, type
IIA and type IIB, and type IB and type I$^{\prime}$. Remarkably,
we will find as a direct consequence of the T-duality
transformations that the tachyonic thermal instabilities arising
in all previous attempts to formulate an equilibrium
supersymmetric string statistical mechanics in the canonical
ensemble are simply {\em absent}.

\vskip 0.1in The earliest works on the statistical mechanics of
strings focussed on the density of states function for the
superconformal field theory (SCFT) on a strip
\cite{hagedorn,bow,rab}, a calculation that enters into the
one-loop vacuum amplitude of an open string theory. The
degeneracy, $P(N)$, at level $N$, with masses scaling as $\left
(N/\alpha^{\prime}\right )^{1/2}$, grows rapidly with increasing
level in any 2d SCFT. Thus, the asymptotic behavior of the density
of states function at high oscillator levels displays an
exponential divergence characterized by the total central charge,
$c$, of the SCFT \cite{hagedorn,bow,rab}. This result follows from
the famous Hardy-Ramanujan formula \cite{hr}:
\begin{equation}
\prod_{k=1}^{\infty} (1-q^k)^{-c} = \sum_N P(N;c) ~ q^N , \quad
\lim_{N \to \infty} P(N;c) = N^{-(c+3)/4} \exp \left [ 2 \pi (c
N/6)^{1/2} \right ] \quad , \label{eq:density}
\end{equation}
where $q$$=$$e^{-2\pi t}$, and $t$ is the edge-length of the
strip. The coefficient in the exponential has become known in the
string theory literature as the Hagedorn temperature associated
with the superconformal field theory of central charge $c$:
\begin{equation}
T_H = 1/2 \pi (c/6)^{1/2} \alpha^{\prime 1/2} \quad .
\label{eq:haged}
\end{equation}
While this is an interesting property of the central charge of the
SCFT including the contribution from both worldsheet bosonic, and
fermionic, modes, notice that this argument has not take into
account the physical state conditions of the supersymmetric {\em
string} theory: these are couched in the form of a projection on
the Hilbert space of the SCFT. The partition function of the SCFT
on a strip is much more than the sum over an infinite set of free
field oscillator contributions of total central charge $c$
accounted for by the Hardy-Ramanujan formula given above: it takes
the form of a sum over integer powers of the Jacobi theta
functions, and each such term contributes to the partition
function with a {\em phase} determined by the physical state
projection on the Hilbert space of the SCFT. The choice of phases,
which can always be interpreted in terms of the choice of {\em
spin structure} for the worldsheet fermions on the annulus, and
which are modified on the nonorientable Riemann surfaces arising
in the presence of an orientifold plane in the embedding target
spacetime as shown in sections 3 and 4, is therefore heavily
constrained by both worldsheet, and {\em target space},
symmetries.

\vskip 0.1in As a consequence of the phases in the partition
function on the strip, an exponential divergence in the
ultraviolet asymptotic density of states function for a SCFT of
given central charge $c$ does {\em not} necessarily signify a
divergence in the vacuum energy density for the finite temperature
string vacuum. A generic property of all supersymmetric string
amplitudes, whether unoriented open and closed, or orientable
closed \cite{gsw,poltorus,typeI,polchinskibook}, is the fact that
worldsheet divergences originating in the ultraviolet asymptotics
of the SCFT can always be mapped to an equivalent divergence of
infrared origin: in the open and closed strings, this is a
consequence of open-closed channel duality which maps the
ultraviolet behavior of the closed string sector to the infrared
of the open string sector, and vice versa. For the orientable
closed superstrings: orientable type IIA, IIB, and heterotic, this
property follows as a consequence of the invariance of closed
string amplitudes under the modular group of the Riemann surface.
Modular invariance can be invoked to excise the domain of the
integral over moduli that would be dominated by contributions from
the deep ultraviolet regime of the closed string spectrum. In
other words, although the Hardy-Ramanujan exponential divergence
in the ultraviolet asymptotics of the free field oscillator sums
in the 2d SCFT is a fact, it is {\em not} necessarily indicative
of a divergence in the energy density of the string thermal
vacuum: the important question is whether there are any
uncancelled {\em infrared} divergences in the string vacuum
amplitude? This requires analysis of the infrared asymptotics of
the vacuum amplitude, separating the individual contributions from
the tachyonic, and massless, modes that dominate the infrared
behavior of the amplitudes. These correspond, respectively, to the
relevant, and marginal, operators in the physical Hilbert space of
the SCFT.

\vskip 0.1in The reader will be familiar with the famous GSO
projection which eliminates the zero temperature tachyon from the
physical spectrum of the supersymmetric string theories
\cite{polchinskibook}. The thermal physical state conditions we
will invoke in what follows achieve a similar projection for all
of the {\em tachyonic} thermal modes in the worldsheet SCFT. With
all of the thermal tachyons projected out of the physical Hilbert
space of the SCFT, and upon verifying the cancellation of massless
tadpoles in the factorization limit of the vacuum amplitude, we
will have succeeded in eliminating all sources of divergence in
the vacuum energy density. It is one purpose of this paper to
explain that this procedure can be successfully carried out for
each of the six supersymmetric string theories. We will find that
each has a viable thermal vacuum in which we can formulate an
equilibrium string statistical mechanics in the canonical
ensemble. In the case of the type IIA and type IIB superstring
theories, we will find that a viable thermal vacuum {\em requires}
the introduction of Dbranes carrying Ramond-Ramond charge, and
consequently, an open string sector with Yang-Mills gauge fields
in the low energy limit.

\vskip 0.1in Having clarified that the canonical ensemble of
supersymmetric strings is well-defined at all temperatures,
including the temperature regime far above the string mass scale,
$\alpha^{\prime -1/2}$, we will establish several new results in
this paper. First, we show that in all of the string ensembles:
heterotic, or type I and type II, the growth of the vacuum energy
density, $\rho(\beta)$, at high temperatures far beyond the string
scale is only as fast as in a 2d quantum field theory. Thus, as
was conjectured with only limited intuition as far back as 1988 by
Atick and Witten \cite{aw}, there is a dramatic reduction in the
{\em growth} of the vacuum energy density at high temperatures.
More recently \cite{polchinskibook}, Polchinski has shown that the
$T^2$ growth in the vacuum energy density at high temperatures is
a direct consequence of the thermal self-duality of the vacuum
functional in the closed bosonic string theory. We will show in
sections 3.2, and 5.1, respectively, that there exists a precise
analog of this behavior for all six supersymmetric string
theories: in each case, the $T^2$ growth in the vacuum energy
density at high temperatures follows as a simple consequence of
the Euclidean T-duality {\em transformations} that link the
thermal ground states of the supersymmetric string theories in
pairs.

\vskip 0.1in The type IB, and type I$^{\prime}$, unoriented open
and closed string ensembles are especially interesting because of
the close comparison that can be made with the finite temperature
supersymmetric Yang-Mills gauge theory obtained in the low energy
limit, where the contribution to the vacuum amplitude from massive
string modes has been suppressed. In particular, we verify in
section 4.3 that the characteristic, stringy, $T^2$ high
temperature growth described above is compatible with the much
faster $T^{10}$ growth at temperatures far below the string scale,
as expected by comparison with finite temperature quantum field
theory. The fact that the energy density grows much more {\em
slowly} at high temperatures beyond the string scale is a hint
indicating a likely thermal phase transition.

\vskip 0.1in Our strategy in section 5 for exposing this phase
transition is as follows: since the string vacuum amplitude has
shown no sign of a discontinuity, or non-analyticity, as a
function of temperature we must look at a different amplitude, or
correlation function, as a plausible order parameter for a thermal
phase transition. A natural choice suggested by the correspondence
in the low energy limit to a finite temperature Yang-Mills gauge
theory, would be the string theory analog of the expectation value
of a timelike Wilson-Polyakov-Susskind loop wrapping the Euclidean
time direction, namely, the change in the free energy in the
thermal vacuum due to the introduction of an external heavy quark,
generally taken to be the order parameter for the deconfinement
phase transition in finite temperature gauge theory
\cite{svet,witads}. We should note that this quantity is extremely
sensitive to infrared divergences in finite temperature gauge
theory, necessitating ingenious techniques for a clear-cut study
of the order parameter in both the lattice, or dual confinement
model, approaches \cite{karsch}. It is generally considered easier
in 4d gauge theory to extract the desired result from a
computation of the {\em pair correlator} of Polyakov-Susskind
loops which yields the static heavy quark-antiquark potential in
the thermal vacuum.

\vskip 0.1in In string theory, it turns out that the Polyakov path
integral summing surfaces with the topology of an annulus and with
boundaries mapped to a pair of fixed curves, ${\cal C}_1$, ${\cal
C}_2$, in the embedding target spacetime, wrapping the Euclidean
time coordinate, and with fixed spatial separation, $R$, can also
be computed from first principles using Riemann surface
methodology. This computation is an {\em extension} of the
one-loop vacuum amplitude computation due to Polchinski
\cite{poltorus}. The amplitude can be interpreted as an off-shell
closed string tree propagator, and the result in closed bosonic
string theory, but only in the limit that the macroscopic
boundaries, ${\cal C}_1$, ${\cal C}_2$, were point-like, was first
obtained by Cohen, Moore, Nelson, and Polchinski \cite{cmnp}.
Their 1986 analysis was recently completed by myself in
collaboration with Yujun Chen and Eric Novak \cite{pair}
including, in particular, the limit of {\em large} macroscopic
loop length of interest to us here. The extension to the
macroscopic loop amplitude in the type I and type II string
theories with Dbranes appears in \cite{pairf}.

\vskip 0.1in Thus, in section 5.2, we will calculate the pair
correlator of a pair of Polyakov-Susskind loops wrapping the
Euclidean time coordinate, extracting the low energy gauge theory
limit of the resulting expression where the contribution from
massive string modes has been suppressed. Notice that in the limit
of vanishing spatial separation, $R$ $\to$ $0$, the amplitude will
be dominated by the shortest open strings, namely, the gauge
theory modes in the massless open string spectrum, and {\em the
worldsheet collapses to a single macroscopic
Wilson-Polyakov-Susskind loop wound around the Euclidean time
coordinate}. We can analyze this limit of the expression for its
dependence on temperature. We find clear evidence for a thermal
phase transition in the gauge theory limit at a transition
temperature of order the string mass scale.

\vskip 0.1in The argument is as follows. A Euclidean T-duality
transformation on our expression for the macroscopic loop
amplitude in the type IB string theory conveniently maps it to an
expression for a corresponding amplitude in the type I$^{\prime}$
string theory. This expression will be well-defined in the
temperature regime above $T_C$, and the low energy gauge theory
limit is easily taken as before. Thus, the existence of T-dual
type IB, and type I$^{\prime}$, descriptions of the thermal ground
state enable precise computations to be made in the low energy
gauge theory limit on either side of the phase boundary. The
intuition that a gas of short open strings transitions into a high
temperature long string phase is an old piece of string folklore,
dating to Hagedorn's 1965 paper
\cite{hagedorn,long,turok,englert,micro,sussk,garyp,rab}.

\vskip 0.1in The outline of this paper is as follows. Section 2
gives an annotated overview of some significant previous
developments in the statistical mechanics of strings, with primary
focus on the worldsheet formulation for the canonical ensemble
following Polchinski's analysis of the closed bosonic string
ensemble in \cite{poltorus}. We clarify the precise differences
between previous attempts, and our work. Section 3.1 contains the
first-principles discussion of the tachyon-free heterotic string
ensemble, emphasizing the significance of the Euclidean T-duality
transformation that links the thermal ground states of the two
heterotic string theories, and clarifying the low energy
interpretation of the timelike Wilson line in terms of the axial
gauge, $A_0$ $=$ $1/\beta$, quantization of the finite temperature
gauge theory. In section 3.2, we demonstrating the $T^2$ growth in
the vacuum energy density at temperatures above the string scale,
and also establish the existence of the Kosterlitz-Thouless
duality transition \cite{kt}. In section 4, we introduce the type
I and type II unoriented open and closed string theories with
Dbranes. Beginning with the closed orientable sector of these
theories, we show in section 4.1 that the requirement of modular
invariance is incompatible with the absence of thermal tachyons in
the pure orientable closed string spectrum at temperatures {\em
above} $T_{\rm w=1}$, the temperature at which the first thermal
winding mode crosses the threshold from irrelevance to marginal
relevance. In section 4.2, we review the concept of
renormalization group (RG) flows induced as a consequence of a
relevant operator in the physical Hilbert space of the SCFT, from
the perspective of both the target spacetime and worldsheet. We
quote the result that follows from the 2d RG analysis in the
accompanying paper \cite{relevant}, namely, that the integrable RG
flow is in the direction towards the infrared fixed point
corresponding to the noncompact, supersymmetric type II vacuum at
zero temperature.

\vskip 0.1in Section 4.3 describes the computation of the one-loop
vacuum energy density at finite temperature in the remaining
sectors of the type IB unoriented open and closed string theory.
Remarkably, we find that summing the contributing worldsheets with
annulus, Mobius Strip, and Klein Bottle topologies gives a {\em
vanishing} one-loop vacuum energy density in the thermal type IB
vacuum. This is despite the fact that target space supersymmetry
has been broken by the thermal type IB physical state conditions.
The vanishing of the one-loop vacuum energy density is a direct
consequence of our requiring the absence of a tadpole for an
unphysical Ramond-Ramond state, a consistency condition we must
impose in the thermal vacuum for the same reasons it was imposed
in the supersymmetric type IB vacuum
\cite{typeI,dbrane,polchinskibook,relevant}. It is most intriguing
to find a nonsupersymmetric type I vacuum, with supersymmetry
broken by finite temperature effects, despite a vanishing one-loop
vacuum energy density, and vanishing dilaton tadpole. Section 5.1
demonstrates the $T^2$ growth in the bosonic contributions to the
type IB vacuum energy density, compatible with the $T^{10}$ growth
in the low energy limit when the contributions from massive string
modes is suppressed. Section 5.2 summarizes our evidence for a
thermal phase transition at the string scale in the low energy
gauge theory limit of the type I string theory. Our conclusions,
and discussion of some future directions suggested by our results,
appear in section 6.

\section{An Overview of the Statistical Mechanics of Strings}

\vskip 0.1in In this section, we provide an annotated overview of
some significant developments in the statistical mechanics of
strings. The development of this subject began almost forty years
ago with Hagedorn's tantalizing suggestion in 1965 of a possible
deconfining phase transition characterized by the appearance of a
\lq\lq long string" in bag models for hadronic physics
\cite{hagedorn}. We emphasize that this rough intuition predates
the identification of QCD$_3$ as the correct theory of the strong
interactions. It also predates the development of perturbative
superstring theory as a unified theory encompassing both quantum
gravity and the nonabelian gauge theories. It is interesting to
note that the earliest motivations for the development of a
statistical mechanics of strings arose in models for the
nonperturbative sector of the strong interactions \cite{hagedorn}
and, subsequently, in models for cosmic string dynamics
\cite{turok,englert}. Both are areas of considerable current
research interest actively exploring the interface with String/M
theory \cite{malda,witads,sundborg,cosmic,lower3}.

\vskip 0.1in Our focus in this paper is on the canonical ensemble
of strings at fixed $(\beta, V)$, and its formulation in terms of
the worldsheet formalism of perturbative string theory following
the appearance of Polchinski's first-principles analysis of the
closed bosonic string ensemble in \cite{poltorus}. We will clarify
the precise differences in the extension of this first-principles
approach to the supersymmetric string ensembles given in our work
from some of the more significant attempts made by other
authors.\footnote{Due to limitations of space, and given that most
of the string literature repeats the misconceptions in both
physical interpretation, or calculation, made in the standard
treatments, we are unable to cite all of the relevant papers in
this area.} A complementary review of the literature the reader
may find useful can be found in \cite{rab}. Polchinski gave the
original derivation of the free energy of the canonical ensemble
of free closed bosonic strings starting with the Polyakov path
integral summing connected Riemann surfaces in \cite{poltorus},
correctly incorporating thermal self-duality of the closed bosonic
string theory \cite{polchinskibook}. However, in physically
interpreting the result for the free energy derived in
\cite{poltorus}, he omits to mention that there are low
temperature tachyonic thermal momentum modes in the spectrum {\em
at all temperatures starting from zero} \cite{bosonic}. Thus,
Polchinski's oft-quoted identification of the first winding mode
instability with the onset of a Hagedorn phase transition is
suspect, an inference independently arrived at with the same
incorrect reasoning by both Kogan \cite{kogan}, and by Sathiapalan
\cite{sathia}. As pointed out by us in \cite{bosonic,relevant},
the bosonic string free energy is already ridden with tachyonic
divergences long before the so-called transition temperature has
been reached: there is no self-consistent equilibrium ensemble of
closed bosonic strings \cite{bosonic}. But the methodology
formulated for this special case by Polchinski in \cite{poltorus}
can be extended with success in the case of all six {\em
supersymmetric} string theories, as was pointed out by us in
\cite{decon}, and in this current work.

\vskip 0.1in Returning to our account of relevant previous
developments following the appearance of \cite{poltorus}, a
proposal for the one-loop free energy of the heterotic string
canonical ensemble was subsequently given by O'Brien and Tan
\cite{bt}: the expression contains tachyonic divergences and is
untenable as a proposal for equilibrium statistical mechanics.
Related discussions of Hagedorn divergences, and of duality in the
thermal mode spectrum, appear in the works \cite{ky,bow,follow}.
The calculation in \cite{bt} did not take into account a crucial
consequence of thermal duality in the heterotic string: namely,
that the $E_8$$\times$$E_8$ and ${\rm Spin}(32)/{\rm Z}_2$
supersymmetric ground states at, respectively, the zero, and
infinite, temperature are interchanged under a thermal duality
transformation on the finite temperature vacuum functional. The
expression for the free energy consequently displays tachyonic
thermal instabilities despite the fact that, unlike the closed
bosonic string, the zero temperature spectrum of either heterotic
string theory is tachyon-free. It is well-known that in the
presence of a tachyonic mode the string ground state is no longer
stable but is liable to evolve under renormalization group flow to
a different infrared-stable vacuum. Such a situation needs to be
analyzed on a case-by-case basis to check the properties of the
infrared-stable endpoint of renormalization group flow: {\em a
string background with a tachyonic thermal mode is not in itself
an acceptable setting for a discussion of equilibrium string
thermodynamics}. Thus, a property we ought to require of the
expression for the string vacuum functional at finite temperature
is the {\em absence} of tachyonic thermal modes. As shown in our
work, unlike the closed bosonic string theory which has a zero
temperature tachyon \cite{poltorus}, in the case of all six
supersymmetric string theories it is indeed possible to arrive at
a tachyon-free thermal spectrum.

\vskip 0.1in Start with the heterotic string theory. How do we
achieve a tachyon-free thermal spectrum, while simultaneously
ensuring that the expression for the heterotic string vacuum
functional satisfies the thermal duality relations as well as all
of the worldsheet consistency conditions required of a string
background, namely, super-Weyl and super-diffeomorphism
invariance, including, in particular, invariance under modular
transformations? The answer for the heterotic string \cite{decon}
can, in fact, be deduced by piecing together results already
existent in the string literature. In 1987, Ginsparg \cite{ginine}
solved an analogous technical problem, but without breaking
supersymmetry, when he gave the general solution for the one-loop
vacuum functional of the circle-compactified heterotic string
theory on $R^{9}$$\times$$S^1$, interpolating between the known
results for the supersymmetric 10D $E_8$$\times$$E_8$ and ${\rm
Spin}(32)/{\rm Z}_2$ strings in the limit of infinite and zero
radius. The reader should keep in mind the correspondence between
compactification radius and inverse temperature: $R$
$\leftrightarrow$ $\beta/2\pi$. Notice that the nonabelian gauge
symmetry at generic radius found in \cite{ginine} was only
$SO(16)$$\times$$SO(16)$: this is an inviolable consequence of
T-duality, which interchanges heterotic string ground states with
different nonabelian gauge group at zero and infinite radius. The
interpolation is achieved by turning on a Wilson line in the
nonabelian gauge theory which is radius dependent, using the
Lorentzian self-dual lattice parametrization of generic heterotic
string backgrounds developed in \cite{nsw}. We now recall another
result from the literature, namely, that there is a {\em unique}
heterotic modular invariant choice of spin structures in nine or
ten dimensions that both breaks supersymmetry and also gives a
tachyon-free spectrum. This result is due to multiple groups of
authors \cite{sw,agmv,dh,klt}. The corresponding nonabelian gauge
group is, happily, $SO(16)$$\times$$SO(16)$. Thus, the circle
compactification of this nonsupersymmetric ground state with the
above-mentioned, radius-dependent Wilson line background
interpolates smoothly between supersymmetric $E_8$$\times$$E_8$
and ${\rm Spin}(32)/{\rm Z}_2$ ground states at zero and infinite
radius and a nonsupersymmetric, but tachyon-free,
$SO(16)$$\times$$SO(16)$ heterotic string theory at intermediate
values of the radius. This intriguing example of a {\em
spontaneous} breaking of supersymmetry by continuous variations of
the string background parameters was explored in both analytic,
and numerical, analyses
 \cite{gv,itoyama}, an idea revived recently in \cite{dienes}. I
 should emphasize that none of these
works made any reference to the physical interpretation of their
results in terms of the behavior of the string canonical ensemble
at finite temperature.

\vskip 0.1in Spontaneous supersymmetry breaking via Scherk-Schwarz
orbifold compactification is an old subject in the string theory
literature \cite{rohm,dudas,bergman}, and conjectural mention of a
possible thermal interpretation has been made for certain
nonsupersymmetric, $(-1)^{F}$, ${\rm Z}_2$ orbifolds of type II,
heterotic, or type I vacua, with target spacetime fermions in a
sector with half-integer timelike momentum \cite{dudas,bergman}.
We believe this physical interpretation is {\em incorrect}. First,
as pointed out above, since all of these nonsupersymmetric ground
states are tachyonic, they are clearly unacceptable as candidates
for describing an {\em equilibrium} string thermodynamics valid in
the full temperature range. Secondly, it should be noted that such
${\rm Z}_2$ orbifold compactifications meet a familiar low energy
finite temperature field theory constraint: namely, that massless
spacetime fermions should acquire a tree-level $1/\beta$ dependent
mass relative to the massless bosons \cite{witads}, by introducing
a half-integer shift in the momenta for spacetime fermions
relative to that for spacetime bosons. In contrast, our
prescription for thermal mode number dependent phases described in
section 3.3 achieves the same low energy constraint {\em without}
invoking any half-integer shifts in thermal momentum. In other
words, our prescription for thermal phases holds in the ordinary
circle-compactified theory and is, therefore, consistent with
Lorentz invariance of the ten-dimensional string
vacuum.\footnote{Note that the expression for the one-loop vacuum
energy density given in Eq.\ (3.125) of the first of the
references in \cite{dudas,dienes} describes an unstable
nonsupersymmetric type IB vacuum, with a closed string tachyon
contributing to the torus amplitude. RR tadpole cancellation
occurs precisely as in the supersymmetric ground state. In
contrast with our expression given in Eq.\ (\ref{eq:freeIp}),
supersymmetry is broken in the one-loop vacuum amplitude by
introducing half-integer momenta for spacetime fermionic states,
relative to integer momenta for spacetime bosons, in the torus,
Klein bottle, and Mobius strip graphs. Although these are
interesting string theory alternatives to traditional scenarios
for dynamical supersymmetry breaking, they do not bear any
relation whatsoever to the thermal vacuum, a point clarified by
consideration of the target spacetime low energy physics. I thank
E.\ Dudas for clarification \cite{update}.}

\vskip 0.1in What is the significance of the radius (inverse
temperature) dependent Wilson line background in the low energy
gauge-gravity field theory limit? Recall that the appropriate
gauge choice for the quantization of finite temperature gauge
theory is axial gauge, $A_0$ $=$ $0$, which correctly accounts for
the requisite physical degrees of freedom in the vector potential.
Of course, this argument does not preclude a modified gauge choice
$A_0$ $=$ $\beta^{-1}$, with $A_0$ set equal to a temperature
dependent constant. In string theory, of course, the inverse
temperature both parameterizes the Wilson line, and also
corresponds to the vev of a scalar field, with corresponding
vertex operator $\partial X^0$, where $X^0$ is Euclidean embedding
time. Thus, the canonical ensemble at fixed temperature
corresponds to a particular string background, and we have a
one-parameter family of Wilson line backgrounds described by the
interpolating vacuum energy density at generic $\beta$. We
emphasize that the choice of Wilson line is determined uniquely by
the requirement that the string vacuum functional transform
correctly under a thermal duality transformation, interchanging
supersymmetric $E_8$$\times$$E_8$ and ${\rm Spin}(32)/{\rm Z}_2$
ground states in the zero, and infinite, inverse temperature
limits. But, happily, the Wilson line also results in a thermal
spectrum that is {\em tachyon-free} at all temperatures starting
from zero, shifting the masses of the would-be thermal tachyons
above the massless threshold. Finally, as we will see below, this
choice of Wilson line also results in a modular invariant
expression for the one-loop vacuum functional.

\vskip 0.1in In summary, it is possible to formulate an
equilibrium string thermodynamics whose low energy limit
corresponds to the equilibrium modes of the canonical ensemble in
a finite temperature 10d gauge-gravity quantum field theory. In
this paper, we verify in section 5.1 that the high temperature
growth in the vacuum energy density as inferred from the low
energy limit of our expression, where the contribution of massive
string modes has been suppressed, indeed reproduces the familiar
$T^{10}$ growth of quantum field theory. This strongly holographic
behavior was originally conjectured by Atick and Witten in 1988
\cite{aw}, and shown to be a consequence of thermal duality in the
closed bosonic string ensemble by Polchinski more recently
\cite{polchinskibook}. It is reassuring that this startling
behavior for the canonical ensemble at temperatures above the
string mass scale is nevertheless compatible with the $T^{10}$
growth of finite temperature field theory reproducible in the low
energy limit.

\vskip 0.1in Let us now clarify a technical issue that has been a
major source of confusion in the literature on the string
canonical ensemble: the choice of boundary conditions on the
worldsheet fermions in the one-loop superstring path integral at
finite temperature. This was discussed at length in Atick and
Witten's work \cite{aw}, regretfully, with an incorrect conclusion
that violates modular invariance. As is well-known, it is the norm
in d-dimensional finite temperature field theory, where $d>2$, to
impose antiperiodic boundary conditions on the fermions in the
direction of Euclidean time. However, we should remind the reader
that {\em there is no spin-statistics theorem in two dimensions}
\cite{ckt}. Thus, while it is important that the low energy limit
of the finite temperature string theory exhibit the familiar
properties expected from a finite temperature 10D quantum field
theory, such considerations apply to the {\em target spacetime}
fermions, rather than the Ramond-Neveu-Schwarz (RNS) worldsheet
fermions. In particular, we are free to choose the spin structure
on the worldsheet in any way we please, subject only to the string
theory consistency conditions \cite{ckt}: super-Weyl and
super-diffeomorphism invariance of the Polyakov path integral
\cite{poltorus,zeta}, including, in particular, modular invariance
of the one-loop vacuum functional in the case of an oriented
closed string theory. We will find that modular invariance forces
us to include sectors with both aperiodic and periodic boundary
conditions on the worldsheet fermions in the closed orientable
type II ensemble.

\vskip 0.1in Consider the possibility of introducing supersymmetry
breaking phases in the string path integral that are thermal mode
number dependent, an idea proposed by Atick and Witten in
\cite{aw}. Such phases will, in fact, be present in the
expressions for the one-loop string free energy of the general
type IIA or type IIB thermal vacua with Ramond-Ramond sector
Dbranes described in later sections of this paper, although we
will motivate them from different considerations: by requiring
self-consistency with the physics of the low energy limit, a
finite temperature gauge theory living on the worldvolume of the
Dbranes. Since there is, in fact, no spin-statistics theorem in
two dimensions, the ad-hoc prescription given by Atick and Witten
in section 5.2 of their paper for the choice of thermal mode
number dependent phases has no justification; notice that no
derivation or proof has been offered. Unfortunately, a serious
problem develops if one introduces a phase factor $(-1)^n$, where
$n$ is thermal momentum mode number, in the one-loop closed string
path integral: such a phase violates modular invariance. Ref.\
\cite{aw} present an argument in sec 5.1 that correlates the
(0,0), (n,0), (0,w), and (n,w) soliton sectors with, respectively,
phase factors: $1$, $(-1)^n$, $(-1)^w$, $(-1)^{n+w}$. The
resulting expression in Eq.\ (5.20) of their paper for the free
energy of the type II oriented closed string ensemble, which we
have reproduced for clarity in Eq.\ (\ref{eq:aws}) of section
34.1, violates modular invariance in the (n,0) and (n,w) sectors.
The corrected modular invariant expression for the one-loop free
energy of the pure closed orientable type II string ensemble
appears in the same section.

\vskip 0.1in We will now explain how the choice of supersymmetry
breaking thermal mode number dependent phases in the string path
integral can, instead, be determined by requiring self-consistency
with the physics of the low energy limit of the string amplitude.
Matching to some well-established property of the field theory
expected in the low energy limit of the string theory puts
constraints on the asymptotics of the one-loop string path
integral and, hence, on any unknown phases present in it. Such low
energy constraints can often be invoked to determine unambiguously
all of the, a priori, undetermined phases in the string path
integral, as emphasized by us in \cite{pairf,zeta}. Consider the
tadpole and anomaly free unoriented 10D type IIB open and closed
string theory with 32 D9branes, or its T-dual 10D type IIA vacuum
with 32 D8branes, at finite temperature. It is clear that the
usual cancellations between spacetime fermions and spacetime
bosons, mass level by mass level, due to target spacetime
supersymmetry must not hold except in the zero temperature limit.
We can introduce supersymmetry breaking phase factors, $(-1)^n$,
in the type IB vacuum functional, where $n$ is thermal momentum
mode number, or $(-1)^w$, where $w$ is thermal winding mode
number, in the dual type I$^{\prime}$ vacuum functional
\cite{decon}. As shown in section 4.3, we will take care to insert
an {\em identical} phase factor for the contributions to the
one-loop vacuum functional from worldsheets with each of the
following one-loop topologies, annulus, Mobius strip, and Klein
bottle, in order to preserve at finite temperatures the familiar
mechanism for the cancellation of the tadpole for the unphysical
Ramond-Ramond scalar that holds in the supersymmetric, zero
temperature, ground state. The contribution from pure closed
oriented one-loop graphs with the topology of a torus vanishes, as
is explained in detail in section 4.2, and in
\cite{decon,relevant}. Let us examine the low energy physics in
the thermal type IB vacuum. It is well known that expansion of the
one-loop modular integral as a function of worldsheet modulus,
$q$$=$$e^{-2\pi t}$, in the limit $t$$\to$$\infty$, enables one to
identify the number of spacetime fermions and spacetime bosons at
each mass level in the open string mass spectrum. Recall that in
the supersymmetric zero temperature vacuum, the absence of a
tachyon {\em required} a relative minus sign between the
contributions of the (AP,AP) and (AP,P) spin structure sectors,
where AP and P denote, respectively, aperiodic and periodic
boundary conditions on the worldsheet (RNS) fermions, implying
that there is no contribution to the one-loop vacuum energy at
order $q^{-1/2}$ \cite{polchinskibook,pairf,zeta}. As shown in
section 4.3, we can preserve this property in the finite
temperature type IB vacuum by a judicious choice of phases, for
every value of thermal momentum mode number, $n$, {\em thereby
ensuring the complete absence of thermal tachyons}.

\vskip 0.1in As in the heterotic string, there is a temperature
dependent Wilson line contributing an overall shift to the vacuum
energy in the finite temperature vacuum. At low temperatures, the
dominant contribution to the type IB vacuum energy is from the n=1
thermal momentum mode. The $(-1)^n$ phase multiplying the (AP,AP)
and (AP,P) sectors has been chosen so that there are no longer any
spacetime fermions contributing at the n=1 level. Thus, {\em
spacetime supersymmetry is spontaneously broken by thermal effects
precisely in the manner expected from generic $d$-dimensional,
$d>2$, finite temperature quantum field theory considerations}
\cite{witads}: the massless spacetime bosons of the zero
temperature vacuum do not acquire a tree-level mass, while the
massless spacetime fermions of the zero temperature vacuum have
acquired an $O(1/\beta)$ tree-level mass. Notice that, as
expected, the spacetime bosons will acquire a mass at the next
order in string perturbation theory, which can be computed by
considering the appropriate one-loop two point function of
massless spacetime boson vertex operators.\footnote{We remind the
reader that the one-loop vacuum amplitude in string theory yields
the tree level mass spectrum, where the radiative corrections to
the mass formula have not been taken into account. The radiative
corrections can be obtained by computing the appropriate two-point
function of vertex operators in string loop perturbation theory.}
We should emphasize that the remaining relative phases in the
expression for the free energy, which is given in Eq.\
(\ref{eq:freeIp}), and the dependence on $N$, the number of
D9branes, are uniquely determined by the requirement of tadpole
cancellation for the unphysical RR scalar in the absence of target
space supersymmetry. Our choice of phases thus ensures that the
usual argument for RR scalar tadpole cancellation in the zero
temperature vacuum \cite{polchinskibook} goes thru unchanged in
the finite temperature expression. This is important, since the RR
unphysical scalar tadpole cancellation is BPS physics, having
nothing to do whatsoever with thermal effects on the type IB
vacuum. Interestingly, it can also be verified that the dilaton
tadpole is also absent as a consequence of requiring the absence
of the unphysical RR sector tadpole. Finally, notice that the
expression in Eq.\ (\ref{eq:freeIp}) for the vacuum functional
interpolates smoothly between the supersymmetric zero temperature
limit, with gauge group $O(32)$, and the finite temperature result
with gauge group $O(16)$$\times$$O(16)$. Remarkably, we have found
a nonsupersymmetric type IB vacuum with supersymmetry broken by
thermal effects, but without the appearance of a dilaton tadpole.
Further details, and equations, can be found in section 4.3.

\vskip 0.1in We should also note that Atick and Witten \cite{aw}
make an incorrect, and unfortunate, assertion about thermal
duality: it is indeed reasonable that the free energy of any
sensible physical system should be a monotonically increasing
function of temperature, but there is nothing \lq\lq unphysical"
about the generating functional of connected vacuum graphs being
thermal self-dual. The Polyakov path integral yields $W(\beta)$,
{\em not} $F(\beta)$, and $W(\beta)$ is indeed thermal self-dual
in the case of the closed bosonic string theory \cite{poltorus}.
The free energy of the string ensemble is given by
$F(\beta)$$=$$-W(\beta)/\beta$, which is, happily, a monotonically
increasing function of temperature. The authors then go on to
dismiss the significance of the thermal duality properties of the
heterotic string vacuum functional. With the exception of the work
by Brandenberger and Vafa \cite{bv}, the notion of thermal duality
has played an increasingly diminishing role in the discussions of
string thermodynamics that have followed Atick and Witten's
influential paper. There has been some investigation of evidence
for the long string transition in the context of a microcanonical
ensemble \cite{long,bv,micro}. More recently, there have been
investigations of the ensemble of type II strings with Dbrane
sources in both the canonical, and microcanonical, approaches
\cite{abel}, but {\em without} an explicit expression for the
one-loop free energy of the canonical ensemble. Regretfully, and
without exception, these works have instead focussed on what we
have pointed out above is a serious physical misinterpretation,
namely, the idea that the first of the tachyonic winding mode
instabilities might be interpreted as a signal for a Hagedorn
phase transition. We reiterate that this physical interpretation
is anyhow {\em untenable}: the thermal spectrum deduced from the
expressions for the one-loop free energy that appear in all of
these previous works is already replete with tachyonic momentum
modes at all temperatures starting from zero.

\vskip 0.1in Our starting point in this paper is the generating
functional of connected one-loop vacuum string graphs, $W(\beta)$
$\equiv$ ${\rm ln}$ ${\rm Z}(\beta)$, derived from first
principles in the Polyakov path integral formalism following
\cite{polyakov,poltorus,zeta}.\footnote{In earlier papers that
include discussion of the bosonic string ensemble as a pedagogical
toy model \cite{bosonic,decon}, we pointed out that, a priori,
there is a peculiar ambiguity in the Euclidean time prescription
for finite temperature quantum theories that can be illustrated by
a canonical ensemble of strings: unlike point particles, strings
are one-dimensional objects sensitive to the topology of the
embedding space. In physics applications where Lorentz invariance
of the target spacetime is not such a sacrosanct principle it
appears to us, a priori, an open question whether Euclidean time
in the thermal prescription has the topology of a circle, or that
of an interval. However, to avoid the misconception that any of
our results for the canonical ensemble of the Lorentz invariant
10D supersymmetric string theories rests on such ambiguity, we
have removed all mention of the issue in this paper. We make the
standard identification of inverse temperature with the radius of
the Euclidean timelike coordinate: $\beta$$=$$2\pi r_{\rm circ.}$,
which has the topology of a circle. But this identification is
unambiguously mandated by the relationship of a given thermal
background of string theory with Euclidean timelike embedding
coordinate, via Wick rotation, to a corresponding Lorentz
invariant background of string theory. It should be noted that a
spontaneous violation of Lorentz invariance can indeed occur in
string/M theory, but this is only in a string vacuum with
nontrivial background field, such as an antisymmetric two-form
potential $B_{\mu\nu}$ \cite{witsei}. To avoid confusion, no such
background fields will be introduced in the Lorentz invariant 10D
flat spacetime string theory backgrounds whose finite temperature
behavior is discussed in this paper.} The vacuum energy density
can, of course, be directly inferred from $W(\beta)$. Let us
recall the basic thermodynamic identities of the canonical
ensemble \cite{pippard}:
\begin{equation}
F= -W/\beta = V \rho , \quad P =
      - \left ( {{\partial F}\over{\partial V}} \right )_T , \quad
      U = T^2 \left ( {{\partial W}\over{\partial T}} \right )_V
, \quad
      S = - \left ( {{\partial F}\over{\partial T}} \right )_V ,
 \quad
      C_V = T \left ( {{\partial S}\over{\partial T}} \right )_V
\quad . \label{eq:freene}
\end{equation}
Note that $W(\beta)$ is an intensive thermodynamic variable
without explicit dependence on the spatial volume. $F$ is the
Helmholtz free energy of the ensemble of free strings, $U$ is the
internal energy, and $\rho$ is the finite temperature effective
potential, or vacuum energy density at finite temperature. $S$ and
$C_V$ are, respectively, the entropy and specific heat of the
thermal ensemble. The pressure of the string ensemble simply
equals the negative of the vacuum energy density, as is true for a
cosmological constant, just as in an ideal fluid with negative
pressure \cite{peebles,pippard}. The enthalpy, $H$$=$$U$$+$$PV$,
the Helmholtz free energy, $F$$=$$U$$-TS$, and the Gibbs function,
also known as the Gibbs free energy, is $G$$=$$U$$-$$TS$$+$$PV$.
As a result of these relations, {\em all} of the thermodynamic
potentials of the string ensemble have been give a simple,
first-principles, formulation in terms of the path integral over
worldsheets. Notice, in particular, that since $P$$=$$-\rho$, the
one-loop contribution to the Gibbs free energy of the string
ensemble vanishes identically! Finally, the reader may wonder why
we omit mention of {\em tree}-level contributions to the vacuum
energy density. In the case of pure closed string theories, or for
the closed string sector of the type I and type I$^{\prime}$ open
and closed string theories, there are none \cite{polchinskibook}:
the underlying reason can be traced to reparameterization
invariance of worldsheets with the topology of a sphere. On the
other hand, in the presence of Dbranes and orientifold planes, one
indeed finds nontrivial tree-level terms in the vacuum energy
density contributed by disk, and crosscap, amplitudes
\cite{polchinskibook,zeta}. However, in the R-R sector
tadpole-free vacuum described in sections 4.3 and 5, the overall
tree-level contribution to the vacuum energy density vanishes
\cite{polchinskibook}.

\vskip 0.1in We have emphasized that many of the significant new
results in this paper: the $T^2$ growth of the string free energy,
the rigorous demonstration of a duality phase transition in the
Kosterlitz-Thouless universality class in the heterotic string,
and the evidence for a novel high temperature long string phase in
the open and closed string theories, are consequences of the
thermal duality transformation properties required of the
expression for the string vacuum functional at finite temperature.
We should mention that a completely different notion of thermal
duality was extensively explored in the early days of string
theory, notably by E.\ Alvarez and collaborators
\cite{ea}.\footnote{These authors also attempt to \lq\lq derive"
the properties of the microcanonical ensemble from those of the
canonical ensemble \cite{ea,micro}, or vice versa, applying them
to models for superstring cosmology. We have already noted that
such attempts are conceptually flawed. A technical reason why the
inverse Laplace transform linking the microcanonical and canonical
closed string ensembles cannot be carried out in closed form is
that the integration over the moduli of the Riemann surfaces
summed in the expression for the closed string free energy cannot
be carried out explicitly.} Duality also makes an appearance in
\cite{kogan,sathia,follow}.\footnote{A much more conjectural
application of {\em thermal self-duality}, based on certain
mathematical identities satisfied by integrals over the
fundamental domain of the modular group of the torus, appears in
recent works by Dienes and Lenneck \cite{newtd}. Since thermal
self-duality is being required of all of the thermodynamic
potentials, the standard relationships in Statistical Mechanics
linking energy and entropy, or energy and specific heat, for
example, no longer hold. Thus, it is unclear how to interpret the
thermal self-dual expressions computed in \cite{newtd} in physical
terms.}

\vskip 0.1in The notion of a holographic principle has played an
active role in recent discussions of quantum gravity and
conjectures for M theory \cite{sussk}. We should emphasize that
the $T^2$ growth of the string free energy at high temperatures
demonstrated in this paper is a much more drastic reduction in the
growth in the number of high temperature degrees of freedom than
is suggested by the holographic principle: $F$ $\propto$ $T^2$ for
the $d$ $=$ $10$ string ensemble, {\em not} simply as fast as an
area, namely, $\propto$ $T^{d-1}$, rather than as a volume,
$\propto$ $T^{d}$, as conjectured in the quantum gravity setting
for a holographic d-dimensional fundamental theory \cite{sussk}.

\section{Canonical Ensemble of Heterotic Closed Strings}

\vskip 0.1in The heterotic closed string theory is possibly the
closest supersymmetric string analog of the closed bosonic string
theory considered by Polchinski in \cite{poltorus}, so let us
begin with this case, as in \cite{decon}. Consider the
ten-dimensional supersymmetric $E_8$$\times$$E_8$ theory at zero
temperature. The $\alpha'$$\to$$0$ low energy field theory limit
is 10D $N$$=$$1$ supergravity coupled to $E_8$$\times$$E_8$
Yang-Mills gauge fields. What happens to the supersymmetric ground
state of this theory at finite temperature? Assuming that a stable
thermal ensemble exists, the finite temperature heterotic ground
state with nine noncompact spatial dimensions is expected to be
tachyon-free, while breaking supersymmetry. Moreover, consistency
with the low energy limit, which is a finite temperature
gauge-gravity theory, implies that the thermal string spectrum
must contain Matsubara-like thermal momentum modes. But the
thermal spectrum is also likely to contain winding modes as
expected in a closed string theory \cite{poltorus}. Most
importantly, since we are looking for a self-consistent string
ground state with good infrared and ultraviolet behaviour, it is
important that the one-loop vacuum functional preserve the usual
worldsheet symmetries of $(1,0)$ superconformal invariance and
one-loop modular invariance. Finally, all of our considerations
are required to be self-consistent with thermal duality
transformation defined as a Euclidean timelike T-Duality
transformation. Since the action of spatial target space dualities
on the different supersymmetric string theories are extremely
well-established, the finite temperature vacuum functional is
required to interpolate between the following two spacetime
supersymmetric limits: in the $\beta$$\to$$\infty$ limit we
recover the vacuum functional of the supersymmetric
$E_8$$\times$$E_8$ heterotic string, while in the $\beta$$=$$0$
limit we must recover, instead, the vacuum functional of the
supersymmetric $\rm Spin(32)/Z_2$ heterotic string. The reason is
that the $E_8$$\times$$E_8$ and $\rm Spin(32)/Z_2$ heterotic
string theories are related by the Euclidean timelike T-duality
transformation: $\beta_{E_8 \times E_8}$$\to$$\beta_{\rm
Spin(32)/Z_2}$$=$$4\pi^2 \alpha'/\beta_{E_8 \times E_8}$.

\subsection{Axial Gauge and the Euclidean Timelike Wilson Line}

\vskip 0.1in  The generating functional of connected one-loop
vacuum string graphs is given by the expression:
\begin{equation}
W_{\rm het.} = L^9 (4\pi^2 \alpha^{\prime})^{-9/2} \int_{\cal F}
{{d^2 \tau}\over{4\tau_2^2}}
  (2 \pi \tau_2 )^{-9/2} |\eta(\tau)|^{-14} Z_{\rm het.} (\beta)
\quad . \label{eq:het}
\end{equation}
The canonical ensemble of oriented closed strings occupies the
box-regularized spatial volume $L^9$. The thermodynamic limit is
approached as follows: we take the limit
$\alpha^{\prime}$$\to$$0$, $L$$\to$$\infty$, holding the
dimensionless combination, $L^9 (4\pi^2 \alpha^{\prime})^{-9/2}$,
fixed. The function $Z_{\rm het.}(\beta)$ contains the
contributions from the rank (17,1) Lorentzian self-dual lattice
characterizing this particular ground state of the circle
compactified $E_8$$\times$$E_8$ heterotic string. Thus, we wish to
identify a suitable interpolating expression for $W_{\rm het}$
valid at generic values of $\beta$, matching smoothly with the
known vacuum functional of the supersymmetric $E_8$$\times$$E_8$
string theory at zero temperature ($\beta$$=$$\infty$):
\begin{eqnarray}
W_{\rm het.}|_{T=0} =&& L^{10} (4\pi^2 \alpha^{\prime})^{-5}
\int_{\cal F} {{d^2 \tau}\over{4\tau_2^2}}
  (2 \pi \tau_2 )^{-5} \cdot
   {{1}\over{|\eta(\tau)|^{16}}} \cdot \left [
   \left ( {{{\bar{\Theta}}_3}\over{\bar{\eta}}}\right )^4 - \left ( {{{\bar{\Theta}}_4
   }\over{\bar{\eta}}}\right )^4 - \left ( {{ {\bar{\Theta}}_2}\over{\bar{\eta}}}\right )^4
   \right ]
   \cr
   && \quad\quad\quad\quad  \times
\left [ \left ({{\Theta_3}\over{\eta}} \right )^8 + \left (
{{\Theta_4}\over{\eta}} \right )^8 + \left
({{\Theta_2}\over{\eta}} \right )^8 \right ]^2 \quad .
\label{eq:holos}
\end{eqnarray}
Thus, $Z_{\rm het.}(\beta)$ describes the thermal mass spectrum of
$E_8$$\times$$E_8$ strings.

\vskip 0.1in It turns out that the desired result already exists
in the heterotic string literature. The modular invariant
possibilities for the sum over spin structures in the 10d
heterotic string have been classified, both by free fermion and by
orbifold techniques \cite{sw,dh,agmv,klt}, and there is a {\em
unique} nonsupersymmetric and tachyon-free solution with gauge
group $SO(16)$$\times$$SO(16)$. Recall the radius-dependent Wilson
line background described by Ginsparg in \cite{ginine} which
provides the smooth interpolation between the heterotic
$E_8$$\times$$E_8$ and $SO(32)$ theories in nine dimensions. We
have: ${\bf A}$$=$${{2}\over{x}}(1,0^7,-1,0^7)$, $x$$=$$
({{2}\over{\alpha^{\prime}}})^{1/2} r_{\rm circ.}$. Introducing
this background connects smoothly the 9D $SO(16)$$\times$$SO(16)$
string at generic radii with the supersymmetric 10d limit where
the gauge group is enhanced to $E_8$$\times$$E_8$. Note that the
states in the spinor lattices of $SO(16)$$\times$$SO(16)$
correspond to massless vector bosons only in the noncompact limit.
Generically, the $(17,1)$-dimensional heterotic momentum lattice
takes the form $E_8$$\oplus$$E_8$$\oplus$$U$. Here, $U$ is the
$(1,1)$ momentum lattice corresponding to compactification on a
circle of radius $r_{\rm circ.}$$=$$x(\alpha^{\prime}/2)^{1/2}$. A
generic Wilson line corresponds to a lattice boost as follows
\cite{ginine}:
\begin{equation}
({\bf p}; l_L, l_R) \to ({\bf p}'; l^{\prime}_L, l^{\prime}_R) =
({\bf p} + w x {\bf A}; u_L - {\bf p}\cdot {\bf A} -
{{wx}\over{2}} {\bf A} \cdot {\bf A} , u_R - {\bf p}\cdot {\bf A}
- {{wx}\over{2}} {\bf A} \cdot {\bf A} ) \quad . \label{eq:boost}
\end{equation}
${\bf p}$ is a 16-dimensional lattice vector in
$E_8$$\oplus$$E_8$. As shown in \cite{ginine}, the vacuum
functional of the supersymmetric 9d heterotic string, with generic
radius {\em and} generic Wilson line in the compact spatial
direction, can be written in terms of a sum over vectors in the
boosted lattice:
\begin{eqnarray}
W_{\rm SS} (r_{\rm circ.} ; {\bf A}) =&& L^{10}(4\pi^2
\alpha^{\prime})^{-5} \int_{\cal F} {{d^2 \tau}\over{4\tau_2^2}}
  (2 \pi \tau_2 )^{-5} |\eta(\tau)|^{-16}
{{1}\over{8}}  {{1}\over{{\bar{\eta}}^4}}
   ({\bar{\Theta}}_3^4 - {\bar{\Theta}}_4^4 - {\bar{\Theta}}_2^4 )
   \cr
   && \quad\quad\quad \times
\left [ {{1}\over{\eta^{16}}} \sum_{({\bf p}'; l^{\prime}_L ,
l^{\prime}_R)} q^{\half ({\bf p}^{\prime 2} +
  l_L^{\prime 2})} {\bar q}^{\half l_R^{\prime 2}}
\right ] \quad .   \label{eq:hollat}
\end{eqnarray}
$W_{\rm SS}$ describes the supersymmetric heterotic string with
gauge group $SO(16)$$\times$$SO(16)$ at generic radius. The
partition function of the nonsupersymmetric but tachyon-free 9d
string with gauge group $SO(16)$$\times$$SO(16)$ at generic radii
is given by \cite{dh,itoyama,sw,agmv,klt,gv}):
\begin{equation}
Z_{\rm NS} (r_{\rm circ.})= {{1}\over{4}} \left [
({{\Theta_2}\over{\eta}})^8 ({{\Theta_4}\over{\eta}})^8
({{{\bar{\Theta_3}}}\over{\eta}})^4 - ({{\Theta_2}\over{\eta}})^8
({{\Theta_3}\over{\eta}})^8 ({{{\bar{\Theta_4}}}\over{\eta}})^4 -
({{\Theta_3}\over{\eta}})^8 ({{\Theta_4}\over{\eta}})^8
({{{\bar{\Theta_2}}}\over{\eta}})^4 \right ]
  \sum_{n,w} q^{\half {\bf l}_L^2 } {\bar{q}}^{\half {\bf l}_R^2}
\quad . \label{eq:dualh}
\end{equation}
However, by identifying an appropriate interpolating function as
in previous sections and appropriate background field, we can
continuously connect this background to the supersymmetric
$E_8$$\times$$E_8$ string.

\vskip 0.1in Since $x$$=$$ ({{2}\over{\alpha^{\prime}}})^{1/2}
{{\beta}\over{2\pi}}$, from the viewpoint of the low-energy finite
temperature gauge theory the timelike Wilson line is simply
understood as imposing a modified axial gauge condition:
$A^0$$=$${\rm const}$. The dependence of the constant on
background temperature has been chosen to provide a shift in the
mass formula that precisely cancels the contribution from low
temperature $(n,0)$ tachyonic modes. As before, we begin by
identifying an appropriate modular invariant interpolating
function:
\begin{eqnarray}
 Z_{\rm het.} =&& {{1}\over{2}}
\sum_{n,w}
 \left [ ({{\Theta_2}\over{\eta}})^8
({{\Theta_4}\over{\eta}})^8 ({{{\bar{\Theta_3}}}\over{\eta}})^4 -
({{\Theta_2}\over{\eta}})^8 ({{\Theta_3}\over{\eta}})^8
({{{\bar{\Theta_4}}}\over{\eta}})^4 - ({{\Theta_3}\over{\eta}})^8
({{\Theta_4}\over{\eta}})^8 ({{{\bar{\Theta_2}}}\over{\eta}})^4
\right ]
  q^{\half {\bf l}_L^2 } {\bar{q}}^{\half {\bf l}_R^2}
\cr &&   {{1}\over{4}} \sum_{n,w} e^{\pi i (n+2w)}\left [ ( {{
{\bar{\Theta_3}} }\over{\eta}} )^4
-({{{\bar{\Theta_2}}}\over{\eta}})^4-({{{\bar{\Theta_4}}}\over{\eta}})^4
\right ] \left [ ({{\Theta_3}\over{\eta}})^{16} + (
{{\Theta_4}\over{\eta}})^{16} + ({{\Theta_2}\over{\eta}})^{16})
\right ] q^{\half {\bf l}_L^2 } {\bar{q}}^{\half {\bf l}_R^2} \cr
\quad && \quad + {{1}\over{2}} \sum_{n,w} e^{\pi i (n+2w)}
 \{ ({{{\bar{\Theta_3}}}\over{\eta}})^4 ({{\Theta_3}\over{\eta}})^8
 \left [
({{\Theta_4}\over{\eta}})^8 + ({{\Theta_2}\over{\eta}})^8 \right ]
- ({{{\bar{\Theta_2}}}\over{\eta}})^4 ({{\Theta_2}\over{\eta}})^8
\left [ ({{\Theta_3}\over{\eta}})^8 + ({{\Theta_4}\over{\eta}})^8
\right ] \cr && \quad \quad \quad -
({{{\bar{\Theta_4}}}\over{\eta}})^4 ({{\Theta_4}\over{\eta}})^8
\left [ ({{\Theta_2}\over{\eta}})^8 + ({{\Theta_3}\over{\eta}})^8
\right ] \}
  q^{\half {\bf l}_L^2 } {\bar{q}}^{\half {\bf l}_R^2} \quad .
\label{eq:iden}
\end{eqnarray}
As in previous sections, the first term within square brackets has
been chosen as the nonsupersymmetric sum over spin structures for
a {\em chiral} type 0 string. This function appears in the sum
over spin structures for the tachyon-free $SO(16)$$\times$$SO(16)$
string given above. Notice that taking the $x$$\to$$\infty$ limit,
by similar manipulations as in the type II case, yields the
partition function of the supersymmetric 10D $E_8$$\times$$E_8$
string.

\vskip 0.1in Consider accompanying the $SO(17,1)$ transformation
described above with a lattice boost that decreases the size of
the interval \cite{ginine}:
\begin{equation}
 e^{-\alpha_{00}} = {{1}\over{1+|{\bf A}|^2/4}} \quad .
\label{eq:scale}
\end{equation}
This recovers the Spin(32)/Z$_2$ theory compactified on an
interval of size $2/x$, but with Wilson line ${\bf A}$$=$$x{\rm
diag}(1^8,0^8)$ \cite{ginine}. Thus, taking the large radius limit
in the {\em dual} variable, and with dual Wilson line background,
yields instead the spacetime supersymmetric 10D $Spin(32)/Z_2$
heterotic string. It follows that the $E_8$$\times$$E_8$ and
$SO(32)$ heterotic strings share the same tachyon-free finite
temperature ground state with gauge symmetry
$SO(16)$$\times$$SO(16)$. The Kosterlitz-Thouless transformation
at $T_C$$=$$1/2\pi\alpha^{\prime 1/2}$ is a continuous thermal
phase transition in this theory.

\vskip 0.1in The thermal duality transition in this theory is in
the universality class of the Kosterlitz-Thouless transition
 \cite{kt}: namely, the partial
derivatives of the free energy to arbitrary order are analytic
functions of temperature at either side of the phase boundary. The
duality transition interchanges thermal momentum modes of the
$E_8$$\times$$E_8$ theory with winding modes of the $Spin(32)/{\rm
Z}_2$ theory, and vice versa. Note that the vacuum functional, the
Helmholtz and Gibbs free energies, the internal energy, and all
subsequent thermodynamic potentials, are both finite and
tachyon-free.

\subsection{Thermal Duality and the High Temperature Degrees of Freedom}

\vskip 0.1in In the closed bosonic string theory, the generating
functional for connected one-loop vacuum string graphs is
invariant under the thermal self-duality transformation: $W(T)$
$=$ $W(T^2_c/T)$, at the string scale, $T_c$ $=$ $1/2 \pi
\alpha^{\prime 1/2}$. As already noted by Polchinski
\cite{polchinskibook}, we can infer the following thermal duality
relation which holds for both the Helmholtz free energy, $F(T)$
$=$ $-T \cdot W(T)$, and the effective potential, $\rho(T)$ $=$
$-T \cdot W(T)/V$ of the closed bosonic string theory:
\begin{equation}
F(T)  = {{T^2}\over{T_C^2}} F({{T_C^2}\over{T}}) , \quad \quad
\rho(T) = {{T^2}\over{T^2_C }} \rho({{T_C^2}\over{T}}) \quad .
\label{eq:thermi}
\end{equation}
In the case of the heterotic string, we will show that the thermal
duality relation instead relates, respectively, the free energies
of the $E_8$$\times$$E_8$ and ${\rm Spin}(32)/{\rm Z_2}$ theories.
Since we deal with a supersymmetric string theory, it is
convenient to restrict ourselves to the contributions to the
vacuum energy density from target space {\em bosonic} degrees of
freedom alone, as was done in \cite{aps,relevant}:
\begin{equation}
F(T)_{E_8\times E_8}  = {{T^2}\over{T_C^2}}
F({{T_C^2}\over{T}})_{{\rm Spin}(32)/{\rm Z}_2} , \quad \quad
\rho(T)_{E_8 \times E_8}  = {{T^2}\over{T^2_C }}
\rho({{T_C^2}\over{T}})_{\rm Spin (32)/Z_2} \quad .
\label{eq:thermih}
\end{equation}
Consider the high temperature limit of this expression:
\begin{equation}
\lim_{T \to \infty } \rho(T)_{E_8 \times E_8} = \lim_{T \to
\infty} {{T^2}\over{T^2_C }} \rho({{T_C^2}\over{T}})_{\rm Spin
(32)/Z_2}
 =  \lim_{(T_C^2/T) \to 0} {{T^2}\over{T^2_C }}
\rho({{T_C^2}\over{T}})_{\rm Spin(32)/Z_2}
 =  {{T^2}\over{T^2_C }} \rho(0)_{\rm Spin(32)/Z_2}
\quad , \label{eq:thermasy}
\end{equation}
where $\rho(0)$ is the contribution to the cosmological constant,
or vacuum energy density, at zero temperature from target space
bosonic degrees of freedom alone. Note that it is finite. Thus, at
high temperatures, the contribution to the free energy of either
heterotic theory from target space bosonic degrees of freedom
alone grows only as fast as $T^2$. In other words, the growth in
the number of degrees of bosonic freedom at high temperature in
the heterotic string ensemble is only as fast as in a {\em
two-dimensional} field theory. This is significantly slower than
the $T^{10}$ growth of the high temperature degrees of freedom
expected in the ten-dimensional low energy field theory.

\vskip 0.1in Notice that the prefactor, $\rho(0)/T_C^2$, in the
high temperature relation is unambiguous, a consequence of the
normalizability of the generating functional of one-loop vacuum
graphs in string theory \cite{poltorus}. It is also background
dependent: it is computable as a continuously varying function of
the background fields upon compactification to lower spacetime
dimension \cite{gv}. The relation in Eq.\ (\ref{eq:thermasy}) is
unambiguous evidence of the holographic nature of perturbative
heterotic string theory: {\em there is a drastic reduction in the
degrees of freedom in string theory at high temperature, a
conjecture first made in \cite{aw}}.

\vskip 0.1in Starting with the duality invariant expression for
the string effective action functional, we can derive the
thermodynamic potentials of the heterotic string ensemble. The
Helmholtz free energy follows from the definition below Eq.\
(\ref{eq:het}), and is clearly finite at all temperatures, with no
evidence for either divergence or discontinuity. The internal
energy of the heterotic ensemble takes the form:
\begin{eqnarray}
 U \equiv&& - \left ( {{\partial W}\over{\partial \beta }} \right )_V
 \cr
 =&&
 L^9 (4\pi^2 \alpha^{\prime})^{-9/2} \half \int_{\cal F}
{{|d\tau|^2}\over{4\tau_2^2}} (2\pi\tau_2)^{-9/2}
   |\eta(\tau)|^{-16}
{{4\pi \tau_2}\over{\beta}} \sum_{n,w \in {\rm Z} }
  \left ( w^2 x^2 - {{n^2}\over{x^2}}  \right )
\cdot q^{ {{1}\over{2}}{\bf l}_L^2 }
     {\bar q}^{{{1}\over{2}}{\bf l}_R^2 }
\cdot Z_{\rm [SO(16)]^2}  , \cr && \label{eq:term}
\end{eqnarray}
where $Z_{\rm [SO(16)]^2}$ denotes the sums over spin structures
appearing in Eq.\ (\ref{eq:iden}). Notice that $U(\beta)$ vanishes
precisely at the string scale, $T_c$$=$$1/2\pi\alpha^{\prime
1/2}$, $x_c$$=$${\sqrt{2}}$, where the internal energy contributed
by winding sectors cancels that contributed by momentum sectors.
Note also that the internal energy changes sign at $T$ $=$ $T_C$.
Hints of this behaviour are already apparent in the numerical
analysis, with plots, of the one-loop effective potential that
appears in \cite{gv,itoyama}.

\vskip 0.1in It is easy to demonstrate the analyticity of
infinitely many thermodynamic potentials in the vicinity of the
critical point. It is convenient to define:
\begin{eqnarray}
[d \tau ] &&\equiv \half L^9 (4\pi^2 \alpha^{\prime})^{-9/2} \left
[ {{|d\tau|^2}\over{4\pi\tau_2^2}} (2\pi\tau_2)^{-9/2}
   |\eta(\tau)|^{-16}
\cdot Z_{\rm [SO(16)]^2} e^{2\pi i nw \tau_1} \right ] \cr
y(\tau_2;x)
   &&\equiv 2\pi \tau_2 \left ( {{n^2}\over{x^2}} + w^2 x^2
  \right ) \quad .
\label{eq:vars}
\end{eqnarray}
Denoting the $m$th partial derivative with respect to $\beta$ at
fixed volume by $W_{(m)}$, $y_{(m)}$, we note that the higher
derivatives of the vacuum functional take the simple form:
\begin{eqnarray}
W_{(1)} =&& \sum_{n,w} \int_{\cal F} [d\tau] e^{-y} (-y_{(1)})
%, \quad
\cr W_{(2)} =&&
 \sum_{n,w} \int_{\cal F} [d\tau] e^{-y}
(-y_{(2)} + (-y_{(1)})^2 ) \cr
%, \quad
W_{(3)} =&& \sum_{n,w} \int_{\cal F} [d\tau] e^{-y} (-y_{(3)} -
y_{(1)} y_{(2)} +  (-y_{(1)})^3 ) \cr
 \cdots =&& \cdots
\cr W_{(m)} =&& \sum_{n,w} \int_{\cal F} [d\tau] e^{-y} (-y_{(m)}
- \cdots +  (-y_{(1)})^m ) \quad . \label{eq:effm}
%\end{equation}
\end{eqnarray}
Referring back to the definition of $y$, it is easy to see that
both the vacuum functional and, consequently, the full set of
thermodynamic potentials, are analytic in $x$. Notice also that
third and higher derivatives of $y$ are determined by the momentum
modes alone:
\begin{equation}
y_{(m)} = (-1)^m n^2 {{(m+1)! }\over{x^{m+2}}} , \quad m \ge 3
\quad . \label{eq:ders}
\end{equation}
For completeness, we give explicit results for the first few
thermodynamic potentials:
\begin{equation}
F = - {{1}\over{\beta}} W_{(0)} , \quad U = - W_{(1)} , \quad S =
W_{(0)} - \beta W_{(1)} , \quad C_V = \beta^2 W_{(2)} , \cdots
\quad . \label{eq:thermodl}
\end{equation}
The entropy is given by the expression:
\begin{equation}
S = \sum_{n,w} \int_{\cal F} [d\tau] e^{-y}
  \left [ 1 + 4\pi \tau_2 ( - {{n^2}\over{x^2}} + w^2 x^2 )
\right ] \quad , \label{eq:entropy1}
\end{equation}
For the specific heat at constant volume, we have:
\begin{equation}
C_V =  \sum_{n,w} \int_{\cal F} [d\tau] e^{-y}
  \left [   16 \pi^2 \tau_2^2 ( - {{n^2}\over{x^2}} + w^2 x^2 )^2
   - 4\pi \tau_2( 3 {{n^2}\over{x^2}} + w^2 x^2 )
\right ] . \label{eq:spc}
\end{equation}
Explicitly, the Helmholtz free energy takes the form:
\begin{equation}
F (\beta) = - \half {{1}\over{\beta}} L^9 (4\pi^2
\alpha^{\prime})^{-9/2}  \int_{\cal F}
{{|d\tau|^2}\over{4\pi\tau_2^2}} (2\pi\tau_2)^{-9/2}
   |\eta(\tau)|^{-16}
\left [ \sum_{n,w} Z_{\rm [SO(16)]^2} q^{ {{1}\over{2}}{{\alpha'
p_L^2}\over{2}} }
    {\bar q}^{{{1}\over{2}} {{\alpha' p_R^2}\over{2}}}
\right ] \quad , \label{eq:freeee}
\end{equation}
while for the entropy of the heterotic string ensemble, we have
the result:
\begin{eqnarray}
S (\beta) =&& \half L^9 (4\pi^2 \alpha^{\prime})^{-9/2} \int_{\cal
F} {{|d\tau|^2}\over{4\pi\tau_2^2}} {{(2\pi\tau_2)^{-9/2}}\over{
   |\eta(\tau)|^{16}}}
\sum_{n,w} \left [
   1 + 4 \pi \tau_2 ( - {{n^2}\over{x^2}} + w^2 x^2 )
\right ] Z_{\rm [SO(16)]^2} q^{ {{1}\over{2}}{{\alpha'
p_L^2}\over{2}} }
    {\bar q}^{{{1}\over{2}} {{\alpha' p_R^2}\over{2}}} .
    \cr &&
\label{eq:entropy}
\end{eqnarray}
The thermodynamic potentials of the heterotic ensemble are finite
normalizable functions at all temperatures starting from zero. In
summary, the heterotic string ensemble displays a continuous phase
transition at the string scale mapping thermal winding modes of
the $E_8$$\times$$E_8$ theory to thermal momentum modes of the
${\rm Spin}(32)/{\rm Z}_2$ theory, and vice versa, unambiguously
identifying a phase transition of the Kosterlitz-Thouless type at
$T_C$$=$$1/2\pi \alpha^{\prime 1/2}$ \cite{kt,bosonic}.

\section{Type I and Type II Open and Closed String Theories}

The Type I and Type II string theories can have both open and
closed string sectors, and the vacuum can contain Dbranes: sources
for Ramond-Ramond charge, with worldvolume Yang-Mills fields
\cite{polchinskibook,zeta}. It is therefore helpful to consider
them in a unified treatment. We will begin with the pure oriented
closed string sector common to all of these theories.

\subsection{Closed Orientable Sector of the Type II String}

\vskip 0.1in We begin with a discussion of the pure type II
oriented closed string one-loop vacuum functional. A Euclidean
T-duality transformation mapping the thermal IIA vacuum to the
thermal IIB vacuum simply maps IIA winding to IIB momentum modes,
and vice versa. In the absence of a Ramond-Ramond sector, the
expression for the normalized generating functional of connected
one-loop vacuum graphs takes the form:
\begin{equation}
W_{\rm II} = L^{9} (4\pi^2 \alpha^{\prime})^{-9/2} \int_{\cal F}
{{d^2 \tau}\over{4\tau_2^2}}
  (2 \pi \tau_2 )^{-4} |\eta(\tau)|^{-14} Z_{\rm II} (\beta)
\quad ,
\label{eq:typeII}
\end{equation}
where the spatial volume $V$$=$$L^9$, while the dimensionless, or
scaled, spatial volume is $ L^{9} (4\pi^2 \alpha^{\prime})^{-9/2}
$. Notice that in the $\alpha^{\prime}$ $\to$ $0$ limit one can
simultaneously take the size of the \lq\lq box" to infinity while
keeping the rescaled volume fixed. This defines the approach to
the thermodynamic limit. The function ${\rm Z}_{\rm II~
orb.}(\beta)$ is the product of contributions from worldsheet
fermions and bosons, ${\rm Z}_F Z_{\rm B}$, and is required to
smoothly interpolate between finite temperature nonsupersymmetric,
and spacetime supersymmetric zero, and infinite, temperature
limits. The spectrum of thermal modes will be unambiguously
determined by modular invariance. Spacetime supersymmetry breaking
will be implemented by the introduction of {\em phases} in the
interpolating function. Such phases can depend on thermal mode
number. They must be chosen {\em compatible with the requirement
that spacetime supersymmetry is restored in the zero temperature
limit of the interpolating function}.

\vskip 0.1in We should note that thermal mode number dependent
phases in the free energy were first proposed by Atick and Witten
in \cite{aw}. There is, unfortunately, a problem with their
particular implementation of this idea that we must point out at
the outset. Let us refer the reader to section 5.1 of \cite{aw}
where the procedure for introducing thermal phases in the string
path integral was first discussed. The thermal spectrum of the
closed type II and heterotic string theories includes both
momentum and winding modes. Since closed strings can wind an
integer number of times around the timelike direction, such a
solitonic sector of the string path integral {\em could} be
weighted by the factor $(-1)^w$, where $w$ is the winding number:
as emphasized in the Introduction, there is no argument based on
the spin-statistics theorem alone for two-dimensional RNS fermions
that {\em requires} such a phase factor. Notice that a $(-1)^w$
phase factor leaves the modular invariance properties of the
one-loop closed string path integral untouched. But let us first
clarify what has been a source of confusion to many readers of our
earlier work \cite{decon}: what goes wrong with modular invariance
in Ref.\ \cite{aw}'s treatment of the type II closed oriented
string ensemble, and why do we report a different result for this
sector of the type I, or type II, thermal ensemble?

\vskip 0.1in Since there is no spin-statistics theorem in two
dimensions, the argument in section 5.2 of \cite{aw}, thru to the
explicit prescription given in Eq.\ (5.4) for the phase factor in
the general $(n,w)$ sector, for every choice of spin structure for
the RNS fermions, has no justification. {\em Notice that there is
no proof, or derivation, given for such a choice of phases.}
Unfortunately, a serious problem develops if one introduces a
phase $(-1)^n$, where $n$ is thermal momentum, in the one-loop
closed string path integral: such a phase violates modular
invariance. AW present an argument in section 5.1 that correlates
the (0,0), (n,0), (0,w), and (n,w) soliton sectors with,
respectively, phase factors: $1$, $(-1)^n$, $(-1)^w$,
$(-1)^{n+w}$. The result for the free energy of the type II
oriented closed string ensemble given in Eq.\ (5.20) of \cite{aw}
is reproduced here:
\begin{eqnarray}
{{F_1}\over{VT}} \sim&& - (4\pi^2 \alpha^{\prime})^{-9/2} \int_F
d^2 \tau \tau_2^{-6} |\eta(\tau)|^{-24} \sum_{n,w} e^{-S_\beta
(n,w)}
 \cr
 &&\quad \times [ (|\Theta_3 |^8
  + |\Theta_4|^8  + |\Theta_2|^8 )(0,\tau)
  + e^{ \pi i (n+w)  }
 (\Theta_2^4 {\bar{\Theta}}^4_4 + \Theta_4^4
 {\bar{\Theta}}_2^4)(0,\tau)
 \cr
 &&\quad\quad
  - e^{i\pi n} (\Theta_2^4 {\bar{\Theta}}_3^4 + \Theta_3^4
  {\bar{\Theta}}_2^4)(0,\tau)
   - e^{i \pi w} (\Theta_3^4 {\bar{\Theta}}_4^4 + \Theta_4^4 {\bar{\Theta}}_3^4 )(0,\tau) ]
\quad , \label{eq:aws}
\end{eqnarray}
where $S_\beta (n,w)$ is the worldsheet soliton action in the
$(n,w)$ sector of the thermal closed oriented string spectrum. As
will be shown below, the second and third terms inside the square
brackets are {\em not} modular invariant, as a consequence of the
thermal momentum mode number dependence in the phase. For the
heterotic string, similar motivational discussion of spin
structures weighted by such phases is presented in sections 5.3,
5.4 and 5.7 of \cite{aw}, but no explicit result for the free
energy was given. The authors appear not to have noticed the clash
of their prescription for phases with one-loop modular invariance.

\vskip 0.1in To be explicit, let us refer the reader to a standard
textbook treatment of the Poisson resummation of the closed string
partition function with momentum and winding modes in a compact
direction of the target spacetime. Keep in mind that inverse
temperature, $\beta$, corresponds to $2\pi R$ in the textbook
discussion. Introduction of a phase, for example, $(-1)^{n+w}$,
would modify the Poisson resummation, the starting point in Eq.\
8.2.9 of the text \cite{polchinskibook}, as follows:
\begin{equation}
|\eta(\tau)|^{-2} \sum_{n,w=-\infty}^{\infty} e^{i \pi (n+w)}
\cdot
    {\rm exp} \left [ - \pi \tau_2 \left ( {{\alpha' n^2
    }\over{R^2}} + {{w^2 R^2}\over{\alpha'}} \right ) + 2 \pi i n
    w \tau_1 \right ]
  \quad , \label{eq:partf}
\end{equation}
We perform a Poisson resummation of the momentum mode number,
namely, $n$:
\begin{equation}
\sum_{n=-\infty}^{\infty} {\rm exp} \left [ -\pi a n^2 + 2 \pi i
bn \right ] = a^{-1/2} \cdot \sum_{m =-\infty}^{\infty}
    {\rm exp} \left [ - \pi (m - b)^2 /a
    \right ]
  \quad . \label{eq:poiss}
\end{equation}
In the absence of a phase factor, the result will be manifestly
invariant under $\tau$$\to$$\tau+1$, if we also let $m$$\to$$m+w$.
The result would also be invariant under $\tau$$\to$$-1/\tau$, if
we simultaneously let $m$$\to$$-w$, $w$$\to$$m$. So, in the
absence of a phase, the partition function is modular invariant.

\vskip 0.1in Are there any possible generalizations with
nontrivial $(n,w)$ dependent phases? Introduction of the phase
factor $(-1)^w$ inside the summation in Eq.\ (\ref{eq:partf}) does
not alter the Poisson resummation on $n$, giving a modular
invariant result. But a phase factor of the form $(-1)^n$, or
$(-1)^{n+w}$, where $n$ is the thermal momentum mode number, as in
the second and third terms of Atick and Witten's result given in
Eq.\ (\ref{eq:aws}), ruins one-loop modular invariance. Upon
including a $(-1)^{n+w}$ phase, the Poisson resummation results in
the following additional terms in the exponent:
\begin{equation}
2\pi R Z_X (\tau) \cdot  \sum_{m,w=-\infty}^{\infty} e^{i \pi w}
\cdot
    {\rm exp} \left [ - \pi R^2  \left \{ |m - w \tau|^2 + (w\tau_1 - m + \quarter ) \right \} /\alpha' \tau_2
    \right ]
  \quad . \label{eq:rpoiss}
\end{equation}
Clearly, the extra terms {\em cannot} be absorbed in a shift on
the summation variable $m$, or $w$. Nor can they be absorbed in
the corresponding modular transform of the products of Jacobi
theta functions arising from the RNS worldsheet fermions, since
these are $w$ independent. This is the basis for our claim that,
by contrast, the expression for the one-loop vacuum functional we
will give below not only satisfies modular invariance, but is also
the {\em unique} acceptable solution.

\vskip 0.1in To reiterate, the supersymmetry breaking thermal mode
number dependent phases must be chosen compatible with the
requirement that spacetime supersymmetry is restored in the zero,
and infinite, temperature limits of the interpolating vacuum
functional of the type IIA (IIB) string ensemble. The {\em unique}
modular invariant interpolating function satisfying these
requirements is:
\begin{eqnarray}
Z_{\rm II} (\beta) =&& {{1}\over{2}} {{1}\over{\eta{\bar{\eta}}}}
\left [ \sum_{w,n \in {\rm Z} }
   q^{{{1}\over{2}} ({{2 \pi \alpha^{\prime}n }\over{\beta}} + {{w\beta}\over{2\pi\alpha^{\prime}}} )^2 }
     {\bar q}^{{{1}\over{2}} ( {{2 \pi \alpha^{\prime}n }\over{\beta}} - {{w\beta}\over{2\pi\alpha^{\prime}}})^2 }
 \right ]
\{ (|\Theta_3 |^8
  + |\Theta_4|^8  + |\Theta_2|^8 ) \cr
  \quad&&\quad\quad + e^{ \pi i w   }
 [ (\Theta_2^4 {\bar{\Theta}}^4_4 + \Theta_4^4 {\bar{\Theta}}_2^4)
  - (\Theta_3^4 {\bar{\Theta}}_4^4 + \Theta_4^4 {\bar{\Theta}}_3^4
   + \Theta_3^4 {\bar{\Theta}}_2^4 + \Theta_2^4 {\bar{\Theta}}_3^4 )
   ] \}
\quad .
\label{eq:bosod}
\end{eqnarray}
The world-sheet fermions have been conveniently complexified into
left- and right-moving Weyl fermions. As in the superstring, the
spin structures for all ten left- and right-moving fermions,
$\psi^{\mu}$, ${\bar{\psi}}^{\mu}$, $\mu$ $=$ $0$, $\cdots$, $9$,
are determined by those for the world-sheet gravitino associated
with left- and right-moving N=1 superconformal invariances.

\vskip 0.1in To understand our result for a modular invariant
interpolating function that also captures the correct thermal
duality properties, first recall the expression for ${\rm
Z}_{SS}$--- the zero temperature, spacetime supersymmetric, limit
of our function given by the ordinary GSO projection:
\begin{equation}
 Z_{\rm SS} =
     {{1}\over{4}}  {{1}\over{\eta^4{\bar{\eta}}^4}}
  \left [ (\Theta_3^4 - \Theta_4^4 - \Theta_2^4)
   ({\bar{\Theta}}_3^4 - {\bar{\Theta}}_4^4 - {\bar{\Theta}}_2^4 )
\right ]
 \quad ,
\label{eq:IIs}
\end{equation}
Notice that the first of the relative signs in each round bracket
preserves the tachyon-free condition. The second relative sign
determines whether spacetime supersymmetry is preserved in the
zero temperature spectrum \cite{pairf}. Next, notice that $Z_{SS}$
can be rewritten using theta function identities as follows:
\begin{eqnarray}
 Z_{\rm SS } &&=
\{ [ |\Theta_3 |^8
  + |\Theta_4|^8  + |\Theta_2|^8 ] \cr
  \quad&&\quad\quad +
 [ (\Theta_2^4 {\bar{\Theta}}^4_4 + \Theta_4^4 {\bar{\Theta}}_2^4)
  - (\Theta_3^4 {\bar{\Theta}}_4^4 + \Theta_4^4 {\bar{\Theta}}_3^4
   + \Theta_3^4 {\bar{\Theta}}_2^4 + \Theta_2^4 {\bar{\Theta}}_3^4 ) ]
  \}
 \quad .
\label{eq:IIid}
\end{eqnarray}
Either of the two expressions within square brackets is modular
invariant. The first expression can be recognized as the
nonsupersymmetric sum over spin structures for the type 0 string
\cite{polchinskibook}. Under a thermal duality transformation,
small $\beta_{IIA}$ maps to large
$\beta_{IIB}$$=$$\beta_C^2/\beta_{IIA}$, also interchanging the
identification of momentum and winding modes,
$(n,w)_{IIA}$$\to$$(n'=w,w'=n)_{IIB}$. Formally, the expression
for the vacuum functional of the IIB closed oriented thermal
ensemble will take identical mathematical form to that for the IIA
closed oriented ensemble.

\vskip 0.1in From the modular invariant expression for the vacuum
energy density of the type II string derived in the previous
section, we can conclude that the physical Hilbert space of the
pure closed orientable thermal type II string contains a tachyonic
instability at all temperatures above $T_{w=1}$, the temperature
at which the first of the thermal winding modes turns tachyonic.
To see this, consider the mass formula in the (NS,NS) sector for
world-sheet fermions, with ${\bf l}_L^2 $ $=$ $ {\bf l}_R^2$, and
$N$$=$${\bar{N}}$$=$$0$:
\begin{equation}
({\rm mass})^2_{nw} =  {{2}\over{\alpha^{\prime}}}
 \left [ - 1 +
{{2 \alpha^{\prime}\pi^2 n^2 }\over{\beta^2}} +
{{\beta^2w^2}\over{8\pi^2 \alpha^{\prime} }}  \right ] \quad .
\label{eq:massII}
\end{equation}
This is the only sector that contributes tachyons to the thermal
spectrum of the SCFT. The potentially tachyonic physical states
are the pure momentum and pure winding states, $(n,0)$ and
$(0,w)$, with $N$$=$${\bar{N}}$$=$$0$. As in the closed bosonic
string analysis, we can compute the temperatures below, and
beyond, which these modes become tachyonic in the absence of
oscillator excitations. Each momentum mode, $(\pm n, 0)$, is
tachyonic {\em upto} some critical temperature, $T^2_n$ $=$ $1/2
n^2\pi^2 \alpha^{\prime} $, after which it turns marginal
(massless). Conversely, each winding mode $(0,\pm w)$, is
tachyonic {\em beyond} some critical temperature, $T^2_w$ $=$
$w^2/8\pi^2 \alpha^{\prime}$. {\em Only the pure winding mode
tachyons survive the thermal physical state conditions of the type
II ensemble}. Thus, at temperatures above $T_{\rm w=1}$, we
apparently do not have a viable equilibrium type II oriented
closed string thermodynamics. Fortunately, as is shown in the
accompanying paper \cite{aps,relevant}, the flow of the worldsheet
RG is in a direction towards the noncompact supersymmetric fixed
point corresponding to, respectively, the type IIA, or type IIB,
vacuum, at zero temperature. A direct corollary of this result
will be, as we show in the next section, that the closed
orientable sub-sector of generic type I and type II string
theories with Dbranes does not break supersymmetry at one-loop
order as a consequence of thermal effects: the contribution to the
one-loop finite temperature vacuum energy density from the torus
vanishes.

\subsection{Cancellation of Ramond-Ramond Sector Tadpole}

\vskip 0.1in We move on to a discussion of the unoriented sectors
of the type IB and type I$^{\prime}$ open and closed string
theories. Consider the one-loop free energy, $F(\beta)$, of a gas
of free type IB strings in the presence of the Euclidean timelike
Wilson line. $F$ is obtained from the generating functional for
connected vacuum string graphs, $F(\beta)$$=$$- W(\beta)/\beta$.
The one-loop free energy receives contributions from worldsurfaces
of four different topologies \cite{polchinskibook,zeta}: torus,
annulus, Mobius strip, and Klein bottle. The torus is the sum over
closed oriented worldsheets and the result is, therefore,
identical to that derived in section 2 for the type IIB string
theory, refer to Eqs.\ (\ref{eq:typeII}) and (\ref{eq:bosod}).
Notice that the closed orientable string sector of the type IB
theory does not distinguish between the Dbrane worldvolume and the
bulk spacetime orthogonal to the branes, since the closed oriented
strings live in all ten dimensions of spacetime. Nor does this
sector have any knowledge about the Yang-Mills sector, or of the
Euclidean timelike Wilson line, since the Yang-Mills fields of
type IB arise as open string excitations. Thus, the comments we
have made earlier regarding worldsheet renormalization group (RG)
flows in the type IIA or type IIB vacuum in the absence of a
Ramond-Ramond sector apply as well for the closed orientable
sub-sector of the generic type IB string theory. Namely, although
we begin with an, a priori, non-vanishing contribution from the
torus amplitude to the vacuum energy density, the thermal spectrum
contains a tachyonic winding mode instability at temperatures
above $T_{\rm w=1}$, and the resulting RG flow is in the direction
restoring the IR stable supersymmetric closed string fixed point
vacuum at zero temperature. In other words, the torus contribution
to the vacuum energy density in the generic unoriented open and
closed type IB thermal vacuum always vanishes \cite{decon}, and
supersymmetry is {\em not} broken by thermal effects in this
sector of the theory at one-loop order.

\vskip 0.1in The contribution to the vacuum energy density from
the remaining three worldsheet topologies that contribute at one
loop order is given by the Polyakov path integral summing surfaces
with two boundaries, with a boundary and a crosscap, or with two
crosscaps \cite{polchinskibook}. We require a tachyon-free thermal
spectrum, namely, a fixed line of thermal type IB vacua
parameterized by the inverse temperature $\beta$ so that the
canonical ensemble describes an equilibrium statistical mechanics
of type IB strings. Finally, as in the supersymmetric type IB
vacuum, consistency requires the absence of a tadpole for the
unphysical Ramond-Ramond (R-R) state in the thermal vacuum
\cite{typeI,dbrane,polchinskibook,relevant}.

\vskip 0.1in On the other hand the usual cancellation between
spacetime fermions and spacetime bosons at each mass level due to
target spacetime supersymmetry must not hold except in the zero
temperature limit. This is in precise analogy with the finite
temperature analysis given earlier for closed string theories and
is achieved by introducing a phase factor, $(-1)^n$ in the type
IB, where $n$ is thermal momentum or type I$^{\prime}$, vacuum
functional \cite{aw}. Note that we insert an identical phase for
the contributions to $F(\beta)$ from worldsheets with each of the
three remaining classes of one-loop graphs--- annulus, Mobius
Strip, and Klein Bottle, in order to preserve the form of the RR
scalar tadpole cancellation in the zero temperature ground state.
The result for the vacuum functional at one-loop order takes the
form:
\begin{eqnarray}
W_{\rm IB}(\beta) = && L^9
(4\pi^2 \alpha^{\prime})^{-9/2} \int_0^{\infty} {{dt}\over{2t}}
  {{(2\pi t )^{-9/2}}\over{
  \eta(it)^{8}}} \sum_{n \in {\rm Z} }
[ ~  N^2 ( Z_{\rm [0]} - e^{\i \pi n } Z_{\rm [1]}) \cr \quad&&
\quad +  2^{10}\left ( Z_{\rm [0]} - e^{\i \pi n } Z_{\rm [1]}
\right ) ~ - ~ 2^6 N \left ( 1 - e^{\i \pi n } \right ) Z_{\rm
[2]} ~ ] \times q^{4\alpha^{\prime}\pi^2 n^2 /\beta^2} . \cr
 && \label{eq:freeIp}
\end{eqnarray}
The thermal modes of the type IB theory correspond to a
Matsubara-like frequency spectrum with timelike momentum spectrum:
$p_n$$=$$2n\pi/\beta$, where $n$$\in$${\rm Z}$. The functions
$Z_{[0]}$, $Z_{[1]}$, and $Z_{[2]}$ are defined as follows:
\begin{eqnarray}
Z_{\rm [0]} =&& ({{\Theta_{00}(it;0)}\over{\eta(it)}})^4 -
({{\Theta_{01}(it;0)}\over{\eta(it)}})^4 \cr Z_{\rm [1]} =&&
({{\Theta_{10}(it;0)}\over{\eta(it)}})^4 -
({{\Theta_{11}(it;0)}\over{\eta(it)}})^4 \cr Z_{\rm [2]} =&& \left
({{\Theta_{01}(it;0)\Theta_{10}(it;0)}\over{\eta(it)\Theta_{00}(it)}}
\right )^4  \quad ,
 \label{eq:relats}
\end{eqnarray}
where the (00), (10), (01), and (11), denote, respectively,
(AP,AP), (P,AP), (AP,P), and (P,P), boundary conditions on
worldsheet fermions along the two cycles of the strip. In this
expression, P and AP denote the periodic and antiperiodic boundary
condition on fermions.

\vskip 0.1in Let us explain our choice of phases in $W_{\rm IB}$
in more detail. As is well known, an asymptotic expansion of the
Jacobi theta functions in powers of $q$$=$$e^{-2\pi t}$ enables
one to identify the number of spacetime fermions and spacetime
bosons at each mass level in the open string spectrum. Recall that
in the supersymmetric zero temperature vacuum, the absence of a
tachyon {\em required} a relative sign between the contributions
of the (AP,AP) and (AP,P) spin structure sectors, denoted (00) and
(01) above, implying that there was no contribution to the
one-loop free energy at $O(q^{-1/2})$
\cite{polchinskibook,pairf,zeta}. We will preserve this property
in the finite temperature vacuum, and for every value of $n$, thus
ensuring the complete absence of thermal tachyons. This is the
reason (AP,AP) and (AP,P) spin structure contributions have been
grouped together in the function $Z_{[0]}$ without any {\em
relative} thermal mode number dependent phase.

\vskip 0.1in The next order in the asymptotic expansion, $O(q^0)$,
gives the massless open string spectrum in the zero temperature
vacuum. The Wilson line contributes an overall shift to the vacuum
energy of the finite temperature vacuum. At low temperatures, the
dominant contribution to the vacuum energy is from the n=1 thermal
momentum mode. The $(-1)^n$ phase multiplying the (10) and (11)
sectors has been chosen so that there are no longer any spacetime
fermions contributing at the n=1 level. Thus, spacetime
supersymmetry is spontaneously broken by thermal effects precisely
in the manner expected from generic $d$-dimensional, $d>2$, finite
temperature quantum field theory considerations: the massless
spacetime bosons of the zero temperature vacuum do not acquire a
tree-level mass, while the massless spacetime fermions of the zero
temperature vacuum have acquired an $O(1/\beta)$ tree-level mass.
Notice that, as expected, the spacetime bosons will acquire a mass
at the next order in string perturbation theory, which can be
computed by considering the appropriate one-loop two point
function of massless spacetime boson vertex operators.\footnote{We
remind the reader that the one-loop vacuum amplitude in string
theory yields the tree level mass spectrum, where the radiative
corrections to the mass formula have not been taken into account.
The radiative corrections can be obtained by computing the
appropriate two-point function of vertex operators in string loop
perturbation theory.} We should emphasize that the remaining
relative phases in Eq.\ (\ref{eq:freeIp}), and the dependence on
$N$, the number of D9branes, are uniquely determined by the
requirement of tadpole cancellation for the unphysical R-R state
even in the absence of target space supersymmetry. Our choice of
phases ensures that the usual argument for the cancellation of the
tadpole for the unphysical R-R state in the zero temperature
vacuum \cite{typeI,dbrane,polchinskibook,relevant} goes thru
unchanged in the finite temperature expression. This is important,
since an uncancelled tadpole for an unphysical R-R state cannot be
removed by an adjustment of the background fields: this would be a
genuine inconsistency of the type IB vacuum, whether at zero
temperature, or at finite temperature. Interestingly, it can be
verified that the dilaton tadpole is also absent in the thermal
vacuum as a consequence of our requiring the absence of the R-R
sector tadpole. Remarkably, we have found a {\em
nonsupersymmetric} type IB vacuum with supersymmetry broken by
thermal effects, but without the appearance of a dilaton tadpole.

\vskip 0.1in A further check of self-consistency with the low
energy gauge theory limit is to verify the expected $T^{10}$
growth of the free energy in the low energy field theory limit. At
low temperatures far below the string scale we expect not to
excite any thermal modes beyond the lowest lying field theory
modes, and should therefore recover the $T^{10}$ growth in the
free energy. Thus, as is standard lore in the string
theory-low-energy-field theory correspondence
\cite{polchinskibook,zeta}, we will expose the $t$$\to$$\infty$
asymptotics of the modular integral in Eq.\ (\ref{eq:freeIp}),
keeping only the leading contribution from the massless modes in
the open string spectrum:
\begin{eqnarray}
F = && - \beta^{-1} \lim_{\beta \to \infty} L^9 (4\pi^2
\alpha^{\prime})^{-9/2} \int_0^{\infty} {{dt}\over{2t}}
  (2\pi t )^{-9/2} \sum_{n \in {\rm Z} }
\left [ 1 + O(e^{-2\pi t}) \right ] q^{4 \alpha^{\prime}\pi^2 n^2
/\beta^2} \cr = && - \beta^{-1} \cdot \beta^{-9} \cdot \rho_{\rm
low} \quad ,
 \label{eq:freeIpt}
\end{eqnarray}
where $\rho_{\rm low}$ is a numerical constant independent of
temperature.\footnote{The precise normalization is given in the
recent paper \cite{micron}.} It is reassuring to recover the
expected $T^{10}$ growth characteristic of a ten-dimensional
finite temperature gauge theory.

\vskip 0.1in Notice that the thermal momentum modes in the
expression for the free energy have no winding mode counterparts
because of the absence of self-duality in the type IB spectrum. In
the Euclidean T-dual type I$^{\prime}$ theory, obtained by letting
$\beta$ $\to$ $\beta_C^2/\beta$ in the expressions above, the type
IB thermal momentum modes are mapped to type I$^{\prime}$ thermal
winding modes, each wrapping the T-dual Euclidean time coordinate
$X^{0\prime}$ \cite{decon}. The type IB timelike Wilson line
wrapping $X^0$ is likewise mapped to the T-dual timelike Wilson
line, wrapping the T-dual coordinate $X^{0\prime}$, with saddle
point action $-\beta^2(2\pi t)/4\pi^2 \alpha^{\prime}$. How should
one interpret the existence of the Euclidean T-dual description of
the thermal ground state of unoriented open and closed type IB
strings? We will close by pointing out that its chief utility is
in the clearer understanding it provides of the high temperature
behavior of the low energy gauge theory limit, where the
contributions from massive string modes has been suppressed.

\vskip 0.2in
\section{High Temperature Transition to the Long String Phase}

It is an old piece of string folklore that at high temperatures,
or at high energy densities, a microcanonical ensemble of strings
at fixed energy will transition into a long string phase. In other
words, most of the ensemble energy resides in one, or more, long,
closed strings, surrounded by a sea of short open strings.
Evidence for such a phase transition has, of course, been of great
interest in models proposed for the dynamics of a cosmic string
ensemble \cite{turok,cosmic}. In the discussion that follows in
section 5.1, we will investigate the high temperature growth in
the number of degrees of freedom in the type I string by an
asymptotic estimate of the growth in the number of states at high
level in the open string mass spectrum. That calculation requires
consideration of the $t$ $\to$ $0$ asymptotics of the integrand in
Eq.\ (\ref{eq:freeIp}). We will find a $T^2$ growth in the free
energy at high temperatures for the type IB ensemble in section
5.1, precisely analogous to that obtained for the heterotic string
ensemble in section 3.2. The dramatic change in the growth in the
free energy above the string scale with a much {\em slower} growth
of the free energy at temperatures $T$ $>>$ $T_C$, is our second,
and more convincing, indication of a plausible thermal phase
transition at a temperature of order the string scale. The order
parameter for this transition will be identified in section 5.2.

\subsection{High Temperature Behavior of the Type I String Ensemble}

\vskip 0.25in We begin by verifying the $T^2$ growth in the vacuum
energy density in the thermal type IB vacuum at high temperatures
above the string scale by direct inspection of the ultraviolet
asymptotics of our expression for the one-loop vacuum functional.
As is well-known \cite{polchinskibook,zeta}, if we wish to sample
the asymptotic behavior at high mass levels of the open string
mass spectrum, we need to consider instead the $t$$\to$$0$
asymptotics of the expression for $F_{\rm IB} (\beta)$. This will
yield an asymptotic estimate for the high temperature behavior of
the full string ensemble.

\vskip 0.1in A modular transformation on the argument of the theta
functions, $t$$\to$$1/t$, puts the integrand in a suitable form
for term-by-term expansion in powers of $e^{-1/t}$; this enables
term-by-term evaluation of the modular integral. Isolating the
$t$$\to$$0$ asymptotics in the standard way \cite{polchinskibook},
we obtain a most unexpected result:
\begin{eqnarray}
\lim_{\beta \to 0} F_{\rm IB} (\beta) = && - \lim_{\beta \to 0}
\beta^{-1}L^9 (8\pi^2 \alpha^{\prime})^{-9/2} \int_0^{\infty}
{{dt}\over{2t}}
  {{t^{-1/2}}\over{
  \eta(i/t)^{8}}}
\sum_{n \in {\rm Z} } [ ~  N^2 ( Z_{\rm [0]} - e^{\i \pi n }
Z_{\rm [1]}) \cr \quad&& \quad\quad\quad +  2^{10}\left ( Z_{\rm
[0]} - e^{\i \pi n } Z_{\rm [1]} \right ) ~ - ~ 2^6 N \left ( 1 -
e^{\i \pi n } \right ) Z_{\rm [2]} ~ ]
 \times
q^{4 \alpha^{\prime}\pi^2 n^2 /\beta^2} \cr
 =&& - \beta^{-1} \sum_{n \in {\rm Z} } \left [ \beta^{-1} \cdot \rho_{\rm high} \right ] + \cdots \quad ,
\label{eq:freeIpph}
\end{eqnarray}
where $\rho_{\rm high}$ is a constant independent of
temperature.\footnote{The precise normalization appears in our
recent work \cite{micron}.} Thus, the canonical ensemble of type
IB strings displays a $T^2$ growth in the free energy at high
temperatures, characteristic of a {\em two-dimensional} field
theory!

\vskip 0.1in It is helpful to verify the corresponding scaling
relations for the first few thermodynamic potentials. The internal
energy of the canonical ensemble of type IB strings takes the
form:
\begin{eqnarray}
 U =&& - \left ( {{\partial W}\over{\partial \beta }} \right )_V
 \cr =&&
 L^9
(4\pi^2 \alpha^{\prime})^{-9/2} \int_{0}^{\infty} {{dt}\over{8t}}
e^{ - 8 \pi^3 \alpha^{\prime} t /\beta^2 }{{(2\pi t
)^{-1/2}}\over{
  \eta(i/t)^{8}}} \cdot \sum_{n \in {\rm Z} } Z_{\rm open}
4 \pi t  \left [ -
   {{ 4 \alpha^{\prime} \pi^2 (1+n^2)}\over{\beta^3}}  \right ]
q^{4 \alpha^{\prime}\pi^2 n^2 /\beta^2 }
 , \cr &&
\label{eq:entermn}
\end{eqnarray}
where ${\rm Z}_{\rm open}(i/t)$ is the factor in curly brackets in
the expression in Eq.\ (\ref{eq:freeIp}). It is clear that the
internal energy of the full ensemble also scales as $\beta^{-2}$
at high temperatures. The entropy is given by the expression:
\begin{equation}
S = - \beta^2 \left ( {{\partial F}\over{\partial \beta }}\right
)_V = - \beta^2 \left [ \beta^{-2} W(\beta) - \beta^{-1} \left (
{{ \partial W}\over{\partial \beta }} \right )_V  \right ] \quad ,
\label{eq:entropy1n}
\end{equation}
and we infer that it scales at high temperatures as $\beta^{-1}$.
Finally, since the specific heat at constant volume is given by:
\begin{equation}
C_V = - \beta \left ( {{\partial S} \over{\partial \beta }} \right
)_V \quad , \label{eq:spcn}
\end{equation}
we infer that it also scales as $\beta^{-1}$ at high temperatures.
Corresponding scaling behavior at {\em low} temperatures, in
agreement with the expected result for the low energy field theory
limit can be extracted, as in the previous section, by considering
instead the $t$$\to$$\infty$ asymptotics of these expressions.

\subsection{An Order Parameter for the Long String Phase Transition}

\vskip 0.1in A plausible order parameter signalling a thermal
phase transition in the type I string theory at $T_H$ is suggested
by the correspondence with the low energy finite temperature gauge
theory limit. It is well known that the order parameter signalling
the thermal deconfinement phase transition in a nonabelian gauge
theory at high temperatures is the expectation value of a closed
timelike Wilson loop \cite{svet,karsch,green,witads}. We wish to
investigate the possibility of a thermal deconfinement phase
transition in the type I theory at a temperature of order the
string scale, conjectured to arise in a gas of short open strings,
and characterized qualitatively by long string formation in the
deconfined high temperature phase
\cite{hagedorn,long,aw,malda,witads,sundborg}.

\vskip 0.1in Since the one-loop vacuum energy density in the type
IB thermal vacuum displays no non-analyticity, or discontinuities,
as a function of temperature, it is natural to look for evidence
in a different string amplitude. A natural choice suggested by the
correspondence in the low energy limit to a finite temperature
Yang-Mills gauge theory, would be the string theory analog of the
expectation value of the Wilson-Polyakov-Susskind loop wrapping
the Euclidean time direction, namely, the change in the free
energy in the thermal vacuum due to the introduction of an
external heavy quark, generally taken to be the order parameter
for the deconfinement phase transition in finite temperature gauge
theory \cite{svet,karsch,witads}.

\vskip 0.1in As mentioned in the introduction, the appropriate
starting point is the Polyakov path integral summing surfaces with
the topology of an annulus and with boundaries mapped to a pair of
fixed curves, ${\cal C}_1$, ${\cal C}_2$, in the embedding target
spacetime, wrapping the Euclidean time coordinate, and with fixed
spatial separation, $R$, can also be computed from first
principles using Riemann surface methodology, an extension of the
one-loop vacuum amplitude computation due to Polchinski
\cite{poltorus}. The amplitude can be interpreted as an off-shell
closed string tree propagator, and the result in closed bosonic
string theory, but only in the limit that the macroscopic
boundaries, ${\cal C}_1$, ${\cal C}_2$, were point-like, was first
obtained by Cohen, Moore, Nelson, and Polchinski \cite{cmnp}, and
extended to include the limit of {\em large} macroscopic loop
lengths of interest here by myself in collaboration with Yujun
Chen and Eric Novak in \cite{pair}. The extension to the
macroscopic loop amplitude in the type I and type II string
theories with Dbranes appears in \cite{pairf}. We will calculate
the pair correlator of a pair of Wilson-Polyakov loops wrapping
the Euclidean time coordinate, extracting the low energy gauge
theory limit of the resulting expression where the contribution
from massive string modes has been suppressed. Notice that in the
limit of vanishing spatial separation, $R$ $\to$ $0$, the
amplitude will be dominated by the shortest open strings, namely,
the gauge theory modes in the massless open string spectrum, and
{\em the worldsheet collapses to a single macroscopic
Wilson-Polyakov-Susskind loop wound around the Euclidean time
coordinate}. Thus, we have a potentially straightforward route in
string theory to extract the standard order parameter \cite{svet}
of the thermal deconfinement transition in the low energy gauge
theory limit. We will analyze this limit of our result for its
dependence on temperature.

\vskip 0.1in Consider the pair correlation function of a pair of
Polyakov-Susskind loops lying within the worldvolume of the
D9branes, and with fixed spatial separation $R$ in a direction
transverse to compactified Euclidean time, $X^0$. Recall that the
boundaries of the worldsheet are the closed \lq\lq
world-histories" of the open string endpoint, which couples to the
gauge fields living on the worldvolume of the Dbranes. The
endpoint state is in the fundamental representation of the gauge
group. Thus, when its closed worldline is constrained to coincide
with a closed timelike loop in the embedding spacetime, the
resulting string amplitude has a precise correspondence in the low
energy limit to the correlation function of two closed timelike
loops representing the spacetime histories of a pair of static,
infinitely massive, quarks with fixed spatial separation. Since we
wish to probe the high temperature behavior of the low energy
gauge theory limit, we should use the Euclidean T-dual type
I$^{\prime}$ description of the thermal vacuum.

\vskip 0.1in The result for the pair correlator of temporal
Wilson-Polyakov loops in the Euclidean T-dual type I$^{\prime}$
vacuum, ${\cal W}^{(2)}_{\rm I^{\prime}}$, derived from first
principles from an extension of the ordinary Polyakov path
integral in the references \cite{cmnp,pair,pairf,zeta}, takes the
remarkably simple form \cite{decon}:
\begin{eqnarray}
{\cal W}^{(2)}_{\rm I^{\prime}} (R,\beta)  =&& \lim_{\beta \to 0}
\int_0^{\infty} dt {{e^{- R^2 t/2\pi \alpha^{\prime}
}}\over{\eta(it)^{8}}}
       \sum_{n\in {\rm Z}} q^{n^2 \beta^2 / 4 \pi^2 \alpha^{\prime} } \cr \quad && \quad \times [~
({{\Theta_{00}(it;0)}\over{\eta(it)}})^4
 -  ({{\Theta_{10}(it;0)}\over{\eta(it)}})^4  \cr
\quad &&\quad - e^{\i \pi n} \{
({{\Theta_{01}(it;0)}\over{\eta(it)}})^4 -
    ({{\Theta_{11}(it;0)}\over{\eta(it)}})^4 \} ~] .
\label{eq:pairc}
\end{eqnarray}
The summation variable, $n$, labels closed string winding modes,
each of which wraps around the Euclidean timelike coordinate
$X^{0\prime}$. The static heavy quark potential can be extracted
as follows. We set ${\cal W}^{(2)}_{\rm I^{\prime}}
$$=$$\lim_{s \to\infty} \int_{-s}^{+s} d s
V[R,\beta]$, inverting this relation to express $V[R,\beta]$ as an
integral over the modular parameter $t$
\cite{dkps,polchinskibook}. The variable $s$ parameterizes proper
time for the pointlike infinitely massive static quarks. Consider
the $q$ expansion of the integrand valid for $t$$\to$$\infty$,
where the shortest open strings dominate the modular integral.
Retaining the leading terms in the $q$ expansion and performing
explicit term-by-term integration over the worldsheet modulus,
$t$, \cite{pair,pairf,zeta}, isolates the following interaction
potential \cite{decon}:
\begin{eqnarray}
V(R,\beta)|_{\beta << \beta_C } =&&  (8\pi^2
\alpha^{\prime})^{-1/2} \int_0^{\infty} dt e^{- R^2 t/2\pi
\alpha^{\prime} } t^{1/2} \quad
 \sum_{n\in {\rm Z}} ( 16 - 16 e^{i \pi n } ) q^{n^2 \beta^2
/4\alpha^{\prime} \pi^2 } + \cdots \cr \quad \simeq &&
{{1}\over{R^3(1+ {{
  \beta^2 }\over{R^2}} )^{3/2} }} + \cdots
\label{eq:static}
\end{eqnarray}
where we have dropped all but the contribution from the $n$$=$$1$
thermal mode in the last step. At high temperatures, with $\beta$
$<<$ $\beta_C$, we can expand the denominator in a power series.
The leading temperature dependent correction to the inverse cubic
power law is, therefore, $O(\beta^2 /R^5)$. At low temperatures,
we must instead use the thermal dual type IB description in order
to extract the low energy gauge theory. We have:
\begin{eqnarray}
{\cal W}^{(2)}_{\rm IB} (R,\beta)  =&& \lim_{\beta \to \infty}
\int_0^{\infty} dt {{e^{- R^2 t/2\pi \alpha^{\prime}
}}\over{\eta(it)^{8}}}
       \sum_{n\in {\rm Z}} q^{4 \pi^2 \alpha^{\prime} n^2 / \beta^2 } \cr \quad && \quad \times [~
({{\Theta_{00}(it;0)}\over{\eta(it)}})^4
 -  ({{\Theta_{10}(it;0)}\over{\eta(it)}})^4  \cr
\quad &&\quad - e^{\i \pi n} \{
({{\Theta_{01}(it;0)}\over{\eta(it)}})^4 -
    ({{\Theta_{11}(it;0)}\over{\eta(it)}})^4 \} ~] \quad ,
\label{eq:pairci}
\end{eqnarray}
from which we can infer:
\begin{eqnarray}
V(R,\beta)|_{\beta >> \beta_C} =&&  (8\pi^2
\alpha^{\prime})^{-1/2} \int_0^{\infty} dt e^{- R^2 t/2\pi
\alpha^{\prime} } t^{1/2} \cr \quad &&\quad \times \sum_{n\in {\rm
Z}} ( 16 - 16 e^{i \pi n } ) q^{4\alpha^{\prime} \pi^2 n^2
/\beta^2} + \cdots \cr \quad \simeq&& {{1}\over{R^3(1+ {{
  16 \pi^4 \alpha^{\prime 2} }\over{R^2 \beta^2}} )^{3/2} }} + \cdots
\label{eq:statici}
\end{eqnarray}
Thus, the low temperature corrections are, instead, a power series
in $\alpha^{\prime 2}/\beta^2 R^2$! In other words, when we use
the appropriate representation of the correlator in the
temperature regime either below, or above, the string scale
transition temperature, we find qualitatively distinct temperature
dependent corrections to the leading inverse cubic power law
static potential. We can interpret this result as evidence for a
phase transition in the low energy gauge theory limit of the type
IB string theory. Remarkably, precise computations can be carried
out on either side of the phase boundary by utilizing,
respectively, the low energy gauge theory limits of a {\em pair}
of thermal dual string theories, type IB and type I$^{\prime}$.

\vskip 0.1in We now make an important observation: since the
expressions given above are analytic as a function of $R$, the
spatial separation of the loops, we can smoothly take the limit
$R$$\to$$0$ where the worldsheet collapses to a single loop. Thus,
the single Wilson-Polyakov-Susskind loop expectation value in the
finite temperature supersymmetric gauge theory transitions from a
$O(1/T^2)$ fall at low temperatures to a $T^2$ growth at high
temperatures above the string scale. From the perspective of
finite temperature $SU(N)$ gauge theory, we would have intuitively
expected an infinite cost in the free energy to produce a single
external quark in the vacuum in the confining
regime.\footnote{Notice that, unlike the expressions for the
string free energy and other thermodynamic potentials in preceding
sections of this paper which are well-defined at all temperatures
starting from zero, the macroscopic loop amplitude in string
theory is not well-defined for Wilson loops at the degenerate
limit points, $T$ $=$ $0$ and $\beta$ $=$ $0$, since the Riemann
surface representation {\em assumes} a worldsheet with closed
boundary loops. Physically, this is simply the statement that the
Polyakov-Susskind loop expectation value is not a good order
parameter for the deconfining phase transition at zero temperature
because the loop is topologically trivial in that limit
\cite{svet}.} However, as explained in \cite{poltorus,dbrane,zeta,
relevant,zeta}, whenever a mass parameter in supergravity, or
gauge theory, is extracted from the low energy field theory limit
of a covariant supersymmetric string theory amplitude, all such
mass terms will be unambiguously normalized in units of the
fundamental string mass scale, $\alpha^{\prime 1/2}$. This is a
consequence of the unique choice of regulator permitted by the
worldsheet gauge symmetries: super-diffeomorphism $\times$
super-Weyl invariance. The dependence on temperature of the
Polyakov-Susskind loop expectation value on either side of the
phase boundary is therefore a very reasonable result from this
perspective.\footnote{The precise normalization appears in our
recent work \cite{micron}.} A more complete discussion of the
numerical significance of this computation will be saved for
future work.

\vskip 0.1in As an aside, it is straightforward to include in this
result the dependence on an external twoform potential or
electromagnetic field. From the previous works
\cite{dkps,pair,pairf,zeta,decon}, the effective string scale in
the presence of an external twoform field strength is further {\em
lowered}. The expression above for the potential can be modified
by the simple replacement: $2\pi\alpha^{\prime
1/2}$$\to$$2\pi\alpha^{\prime 1/2}u$, where ${\cal
F}^{09}$$=$${\rm tanh}^{-1}u$ is the electric field strength.
Notice that the phase transition temperature would, therefore, be
modified by the factor $u^{-1}$ from the zero field prediction. As
mentioned in Footnote 5, we do not wish to confuse the reader by
introducing two-form background fields that lead to a spontaneous
breaking of target space Lorentz invariance, so we will simply
omit the corresponding equations. Related discussion can be found
in \cite{lower3}.

\section{Conclusions}

\vskip 0.1in We have shown that each of the six supersymmetric
string theories: heterotic, type I, and type II, admits stable
thermal backgrounds in which we can formulate an equilibrium
statistical mechanics of strings in the canonical ensemble. We
have explained in the Introduction, and in Section 2, why
preceding attempts \cite{bow,ky,bt,follow,aw,micro,bv,abel,rab} to
formulate an equilibrium statistical mechanics with a tachyon-free
supersymmetric string canonical ensemble have failed: these works
did not correctly incorporate the Euclidean T-duality
transformations linking the thermal vacua of the supersymmetric
string theories in pairs. In particular, the widespread belief
that the canonical ensemble of supersymmetric strings breaks down
beyond the limiting Hagedorn temperature, $T_H$, turns out to be
{\em incorrect}.

\vskip 0.1in In the case of the type IIA and type IIB superstring
theories, we have shown that a viable thermal vacuum {\em
requires} the introduction of Dbranes carrying Ramond-Ramond
charge, and consequently, an open string sector with Yang-Mills
gauge fields in the low energy field theory limit. Our analysis
holds at one-loop order in the perturbation expansion in the
string coupling constant $g_s$. We remind the reader that the one
loop contribution to the vacuum energy density is not accompanied
by explicit powers of $g_s$ and our conclusions should, therefore,
dovetail neatly with any fully nonperturbative string/M theory
analysis that follows in the future. But we have also exhibited
some tantalizing properties such as the absence of the dilaton
tadpole, and a vanishing vacuum energy density, at one-loop order
in the {\em nonsupersymmetric} unoriented open and closed type IB
thermal vacuum with anomaly-free gauge group
$O(16)$$\times$$O(16)$. This is a remarkable result from the
perspective of proposals for spontaneous supersymmetry breaking in
string theory \cite{relevant,dudas}. We believe there is sound
physical motivation here for a more careful consideration of the
string loop corrections to our results in Section 4. Fortunately,
there have been some new advances in the understanding of type II
superstring loop amplitudes at higher genus in recent years
\cite{dhoker}, and we defer further discussion to future work.

\vskip 0.1in  Next, we have shown that the growth of the vacuum
energy density of the string ensemble at high temperatures far
above the string scale is only as fast as that in a {\em
two-dimensional} quantum field theory. This high temperature
growth holds for all six supersymmetric string theories. In the
case of the type I unoriented open and closed superstring where
the low energy gauge theory limit can be readily extracted in
closed analytic form, it was reassuring to find in section 4.3
that our expression for the free energy of the string ensemble
also reproduces the expected $T^{10}$ quantum field theoretic
growth at temperatures much below the string scale, where the
contribution from massive string modes has been suppressed. In
section 3.2, we have established the existence of a duality phase
transition in the Kosterlitz-Thouless universality class mapping
the finite temperature ground state of the $E_8$$\times$$E_8$
heterotic string to its ${\rm Spin}(32)/{\rm Z}_2$ Euclidean
T-dual.

\vskip 0.1in Most importantly, the evidence we have given in
Section 5 in favor of a novel long string phase at high
temperatures in the microcanonical ensemble of short open strings
in the type IB string theory is strongly deserving of further
investigation, especially as motivated by models for the dynamics
of a cosmic string network \cite{turok,long,micro,cosmic}. The
question of interest here is establishing concrete evidence for
the {\em scaling regime}, dominated by one, or more, long winding
mode cosmic strings in a bath of thermal radiation described by
the short loop length limit of a microcanonical ensemble of open
and closed strings. It would also be interesting in this context
to extend our analysis for the anomaly-free type IB vacuum with 32
D9branes to an anomaly-free type IB thermal vacuum with $N$
additional D9brane-anti-D9brane pairs \cite{sugimoto,schw}. We
should note that D-Dbar annihilation has played an important role
in recent proposals for generating a suitable inflationary
potential in models for early universe superstring cosmology
\cite{cosmic}. A further motivation comes from finite temperature
large $N$ supersymmetric gauge theory
\cite{svet,witads,malda,rab,sundborg}. It would be extremely
interesting to explore the low energy gauge theory limit in the
{\em large $N$ limit} of the anomaly-free type IB string theories
with $32$$+$$N$ D9branes and $N$ anti D9branes constructed in
\cite{sugimoto,schw}; the one-loop string vacuum amplitude for
finite $N$ is given explicitly in the first reference.

\vskip 0.1in We will close with mention of two important insights
that apply more broadly to future developments in String/M Theory.
First, we should remind the reader that perturbative string theory
as formulated in the worldsheet formalism is inherently background
dependent: the \lq\lq heat-bath" representing the embedding target
space of fixed spatial volume and inverse temperature is forced
upon us, together with any external background fields
characterizing the spacetime geometry. Thus, we are ordinarily
restricted to the canonical ensemble of statistical mechanics. We
should caution the reader that while an immense, and largely
conjectural, literature exists on proposals for microcanonical
ensembles of weakly-coupled strings \cite{ea,bt,long,bv,micro},
the conceptual basis of these treatments is full of holes. Some of
the pitfalls have been described in \cite{svet,aw}. It should be
kept in mind that, strictly speaking, the microcanonical ensemble
is what is called for when discussing quantum cosmology, or the
statistical mechanics of the Universe \cite{hawk}: the Universe
is, by definition, an isolated closed system, and it is
meaningless to invoke the canonical ensemble of the \lq\lq
fundamental" degrees of freedom. However, there are many simpler
questions in both early Universe cosmology
\cite{turok,bv,giov,abel,rab,cosmic}, and in the high temperature
behavior of low energy supersymmetric gauge theories
\cite{hagedorn,svet,karsch,witads,sundborg,rab}, that are indeed
approachable within the framework of statistical mechanics in the
canonical ensemble, such as our discussion of the long string
transition in Section 5. In this context, it would be interesting
to explore the properties of the microcanonical ensemble of
weakly-coupled type IB-I$^{\prime}$ unoriented strings described
in Section 5 using some of the methodology outlined in
\cite{bt,long,micro,bv,abel,rab}. The key difference here is that
the one-loop free energy of type I strings {\em vanishes} as a
consequence of RR sector tadpole cancellation. Thus, there is no
room for thermal back-reaction and the fixed energy microcanonical
ensemble should be well-defined \cite{micro}.

\vskip 0.05in A second important insight from our analysis of the
thermal vacua of the different supersymmetric string theories is
that the 10D type II superstring backgrounds with trivial
Ramond-Ramond sector, corresponding to pure supergravity theories
with 32 supercharges in the low energy limit, are a somewhat
artificial truncation of the more generic type II superstrings
with Dbranes. These are theories with 16, or fewer, supercharges,
and they have both super-Yang-Mills fields and supergravity in
their low energy limit, in common with generic backgrounds of the
type I and heterotic string theories. Recovering the 10D
backgrounds with 32 supercharges as special limits characterized
by an {\em enhanced supersymmetry} in a theory where the generic
backgrounds contain both supergravity and super-Yang-Mills sectors
and, consequently, 16 or fewer supercharges, is a key element of
the matrix proposal for a fundamental theory of emergent spacetime
under development in \cite{mtheory,relevant,hodge}. We comment
that such a formalism places the embedding spacetime geometry, and
the quantized degrees of freedom within it, on equal footing,
thereby implementing the full spirit of Einstein's vision for a
fundamental theory. It is also a viable starting point for a
discussion of the quantum statistical mechanics of an isolated
closed system like our Universe.

\vspace{0.1in} \noindent{\bf Acknowledgements:} Early developments
in this research project were supported in part by
NSF-PHY-9722394, and reported upon in \cite{decon,bosonic}. I
would like to acknowledge Joe Polchinski for urging me to be more
precise in applying the thermal (Euclidean T-duality) {\em
transformations} linking the thermal vacua of the six Lorentz
invariant 10D supersymmetric string theories. The results were
more recently updated at the Aspen Center for Physics, following
the renormalization group analysis presented in \cite{relevant}. I
am grateful to Hassan Firouzjahi for pointing out a minor, but
embarrassing, typo in earlier treatments of the tachyonic type II
ensemble, and to Scott Thomas for the invitation to present this
work at the {\em Cosmic Acceleration} workshop. I thank S.\ Abel,
Z.\ Bern, K.\ Dienes, E.\ Dudas, H.\ Firouzjahi, G.\ Horowitz, M.\
Gutperle, D.\ Kabat, M.\ Kleban, P.\ Krauss, A.\ LeClair, E.\
Mottola, G.\ Semenoff, S.\ Raby, H.\ Tye, and the seminar
participants at the KITP {\em QCD and String Theory} workshop, for
helpful questions and comments. It is a pleasure to acknowledge
the participants at the Columbia University {\em 4th Northeast
String Cosmology Meeting} for stimulating discussions.

\vskip 0.2in \noindent{\bf Note Added (Sep 2005):} This document 
has far too much material
to be readable, and I have separated the useful 
content into several follow-up papers. The historical account in
section 2 may be useful for some readers, except that it should be noted that
thermal mode number dependent phases in the type II string vacuum
amplitude do not work: it is not possible to spontaneously break
spacetime supersymmetry compatible with modular invariance. The
conclusion that there is no viable type II canonical ensemble in the
absence of a Yang-Mills sector stands. A better discussion of the type
II orientifold is given in follow-up work (hep-th/0506143).


\begin{thebibliography}{99}
\bibitem{poltorus}J. Polchinski,
{\em Evaluation of the One Loop String Path Integral}, Comm. Math.
Phys. {\bf 104} (1986) 37.
\bibitem{aw}J. Atick and E. Witten,
{\em The Hagedorn Transition and the Number of Degrees of Freedom
in String Theory}, Nucl.\ Phys.\ {\bf B310} (1988) 291.
\bibitem{decon} S. Chaudhuri, {\em Deconfinement
and the Hagedorn Transition in Thermal String Theory}, Phys.\
Rev.\ Lett.\ {\bf 86} (2001) 1943, hep-th/0008131.
\bibitem{bosonic}S. Chaudhuri, {\em Finite Temperature Closed
Bosonic Strings: Thermal Duality and the Kosterlitz-Thouless
Transition}, Phys.\ Rev.\ {\bf D65} (2002) 066008, hep-th/0105110.
\bibitem{relevant}S.\ Chaudhuri, {\em Decompactification and
the g--theorem}, hep-th/0408206.
\bibitem{polyakov}A.\ M.\ Polyakov, {\em Quantum Geometry of
Bosonic Strings}, Phys.\ Lett.\ {\bf B103} (1982) 107.
\bibitem{hawk}S.\ W.\ Hawking, {\em Quantum Gravity and Path
Integrals}, Phys.\ Rev.\ {\bf D18} (1978) 1747. {\em Zeta Function
Regularization of Path Integrals}, Comm.\ Math.\ Phys.\ {\bf 55}
(1977) 133.
\bibitem{typeI}J.\ Polchinski and Y.\ Cai, {\em Consistency of Open
Superstring Theories}, Nucl.\ Phys.\ {\bf B296} (1988) 91.
\bibitem{ohta} N.\ Ohta, Phys.\ Rev.\ Lett.\ {\bf 59} (1987) 176.
\bibitem{pippard} A.\ B.\ Pippard, {\em The Elements of
Classical Thermodynamics}, Cambridge.
\bibitem{peebles} P.\ J.\ E.\ Peebles and B.\ Ratra,
{\em The Cosmological Constant and Dark Energy}, Rev.\ Mod.\
Phys.\ {\bf 75} (2003) 559.
\bibitem{hagedorn} R. Hagedorn, Nuovo Cim.\ Suppl.\ 3 (1965) 147.
K. Huang and S. Weinberg, Phys.\ Rev.\ Lett.\ {\bf 25} (1970) 895.
S. Fubini and G. Veneziano, Nuovo Cim.\ {\bf 64A} (1969) 1640. R.
Carlitz, Phys. Rev. {\bf D5} 3231 (1972). S. Frautschi, Phys. Rev.
{\bf D3} 2821 (1971). C.\ Thorn, Phys.\ Lett.\ {\bf 99B} (1981)
458. R.\ Pisarski, Phys.\ Rev.\ {\bf D29} (1984) 1222.
\bibitem{hr}G.\ Hardy and S.\ Ramanujan, Proc.\ Lond.\ Math.\
Soc.\ {\bf 17} (1918) 75.
\bibitem{ea}E.\ Alvarez, {\em Strings at Finite Temperature}, Nucl.\
Phys.\ {\bf B269} (1986). E.\ Alvarez and M.\ A.\ Osorio, {\em
Superstrings at Finite Temperature}, Phys.\ Rev.\ {\bf D36} (1987)
1175.
\bibitem{ky}
K. Kikkawa and M. Yamasaki, {\em Casimir Effects in Superstring
Theories}, Phys.\ Lett.\ {\bf B149} (1984) 357. N. Sakai and I.
Senda, {\em Vacuum Energies of String Compactified on Torus},
Prog.\ Theor. Phys. 75 (1986) 692.
\bibitem{din1}M.\ Dine, P.\ Huet, and N.\ Seiberg, Nucl.\ Phys.\
{\bf B322} (1989) 301.
\bibitem{nair}V.\ P.\ Nair, A.\ Shapere, A.\ Strominger, and F.\ Wilczek,
{\em Compactification of the Twisted Heterotic String Torus},
Nucl.\ Phys.\ {\bf B287} (1987) 402.
\bibitem{busch}T.\ Buscher, Phys.\ Lett.\ {\bf 194B} (1987) 59;
{\bf B201} (1988) 466.
\bibitem{dlp}J.\ Dai, R.\ Leigh, and J.\ Polchinski, Mod.\ Phys.\
Lett.\ {\bf A4} (1989) 2073.
\bibitem{bow}M.
Bowick and L.C.R. Wijewardhana, {\em Superstrings at High
Temperature}, Phys.\ Rev.\ Lett.\ {\bf 54} 2485 (1985). S.\-H.\
Henry Tye, {\em The Limiting Temperature Universe and
Superstrings}, Phys.\ Lett.\ {\bf B158} (1985) 388.
\bibitem{kogan} Ya. I. Kogan, {\em
Vortices on the World-Sheet and String's Critical Dynamics}, JETP
Lett.\ {\bf 45} (1987) 709.
\bibitem{sathia}B. Sathiapalan, {\em
Duality in Statistical Mechanics and String Theory}, Phys.\ Rev.\
Lett.\ {\bf 58} (1987) 414.b
\bibitem{follow}B. Maclain and B. Roth,
Comm.\ Math.\ Phys.\ {\bf 111} (1987) 1184.
\bibitem{bt}K.\ O'Brein and
C.\-I.\ Tan, {\em Modular Invariance of the Thermopartition
Function and Global Phase Structure of the Heterotic String},
Phys.\ Rev.\ {\bf D36} (1987) 1184. N. Deo, S. Jain, and C.-I.
Tan, {\em Strings at High Energy Densities and Complex
Temperature}, Phys.\ Lett.\ {\bf B220} (1989) 125; {\em String
Statistical Mechanics above Hagedorn Energy Densities}, Phys.\
Rev.\ {\bf D40} (1989) 2626.
\bibitem{long}B.\ Sundborg, Nucl.\ Phys.\ {\bf B254} (1985) 583.
P.\ Salomonson and B.\ S.\ Skagerstam, Nucl.\ Phys.\ {\bf B268}
(1986) 349; {\em Physica} {\bf A158} (1989) 499.
\bibitem{turok}D.\ Mitchell and N.\ Turok, Phys.\ Rev.\ Lett.\
{\bf 58} (1987) 1577; {\em Statistical Properties of Cosmic
Strings}, Nucl.\ Phys.\ {\bf B294} (1987) 1138.
\bibitem{englert}F.\ Englert, J.\ Orloff, and T.\ Piran, {\em Fundamental Strings
and Large Scale Structure Formation}, Phys.\ Lett.\ {\bf B212}
(1988) 423.
\bibitem{bv} R. Brandenberger and C. Vafa,
{\em Superstrings in the Early Universe}, Nucl.\ Phys.\ {\bf B316}
(1989) 391.
\bibitem{micro}M. Bowick and S. Giddings, {\em High Temperature Strings}, Nucl.\
Phys.\ {\bf B325} (1989) 631. A. Bytsenko, E. Elizalde, S. D.
Odintsov and S. Zerbini, Nucl.\ Phys.\ {\bf 394} (1993) 423. D.\
Lowe and L. Thorlacius, {\em Hot String Soup}, Phys.\ Rev.\ {\bf
D51} (1995) 665. M.\ C.\ Abdalla, A.\ L.\ Gadelha, and D.\ Nedel,
{\em Closed String Thermal Torus from Thermofield Dynamics},
hep-th/0410068.
\bibitem{kt} J. M. Kosterlitz and D.\ M.\ Thouless,
J.\ Phys.\ {\bf C6} (1973) 1181. J. Cardy, {\em Scaling and
Renormalization in Statistical Physics}, Chap.\ 6 (Cambridge)
1996. C. Itzykson and J. Drouffe, {\em Statistical Field Theory},
Vol.\ I, sec.\ 4.2 (Cambridge).
\bibitem{gsw}M.\ Green, J.\ Schwarz, and E.\ Witten, {\em Superstring
Theory}, in two volumes (Cambridge) 1987.
\bibitem{polchinskibook} J. Polchinski, {\it String Theory}, in
two volumes (Cambridge) 1998. Sec 9.8 contains the demonstration
that the high temperature $T^2$ growth in the vacuum energy
density in the closed bosonic string theory follows as a
consequence of its thermal self-duality.
\bibitem{zeta}S.\ Chaudhuri, {\em The Normalization of
Perturbative String Amplitudes: Weyl Covariance and Zeta Function
Regularization}, electronic pedagogical review article,
hep-th/0409031.
\bibitem{kirzhlinde}
R.\ Jackiw, Phys.\ Rev.\ {\bf D9} (1974) 1686. A. Linde, {\em
Phase Transitions in Gauge Theories and Cosmology}, Rept.\ Prog.\
Phys.\ {\bf 42} (1979) 389.
\bibitem{dinesei} M. Dine and N. Seiberg, {\em Is the Superstring
Weakly Coupled?}, Phys.\ Lett.\ {\bf B162} 299 (1985).
\bibitem{cs}S.\ Chaudhuri and J.\ Schwartz, {\em A Criterion for
Integrably Marginal Operators}, Phys.\ Lett.\ {\bf B219} (1989).
\bibitem{al}I.\ Affleck and A.\ W.\ W.\ Ludwig, {\em Universal Non-Integer Ground State Degeneracy
in Quantum Critical Systems}, Phys.\ Rev.\ Lett.\ {\bf 67} (1991)
161.
\bibitem{aps}J.\ Polchinski, {\em Scale and Conformal Invariance
in String Theory}, Nucl.\ Phys.\ {\bf B303} (1988) 226. A.\ Adams,
J.\ Polchinski, and E.\ Silverstein, {\em Don't Panic! Closed
String Tachyons in ALE Spacetimes}, JHEP 0110 (2001) 029,
hep-th/0108075.  I.\ Kani and C.\ Vafa, {\em Asymptotic Mass
Degeneracies in Conformal Field Theories}, Phys.\ Lett.\ {\bf
B130} (1990) 529. J.\ Harvey, D.\ Kutasov, E.\ Martinec, and G.\
Moore, {\em Localized Tachyons and RG Flows}, hep-th/0111154.
\bibitem{nsw} K.S. Narain, M. Sarmadi, and E. Witten,
{\em A Note on Toroidal Compactification of Heterotic String
Theory}, Nucl.\ Phys.\ {\bf B279} (1987) 369.
\bibitem{ginine} P. Ginsparg, {\em Comment on Toroidal
Compactification of Heterotic Superstrings},
Phys.\ Rev.\ {\bf D35} 648 (1987).
\bibitem{sw}N. Seiberg and E. Witten, {\em Spin Structures in String
Theory}, Nucl.\ Phys.\ {\bf B276} (1986) 272.
\bibitem{ckt} S. Chaudhuri, H. Kawai, and S.-H.H. Tye,
{\em Path Integral Formulation of Strings}, Phys.\ Rev.\ {\bf D36}
1148 (1987). See section 3.
\bibitem{agmv}L. Alvarez-Gaume, P. Ginsparg, G. Moore, and C. Vafa,
{\em An $O(16)$$\times$$O(16)$ Heterotic String},
Phys.\ Lett.\ {\bf B171} (1986) 155.
\bibitem{dh}L. Dixon
and J. Harvey, {\em String Theories in Ten Dimensions
Without Spacetime Supersymmetry},
Nucl.\ Phys.\ {\bf B274} 93 (1986).
\bibitem{klt}H. Kawai, D. Lewellen, and S.-H. H. Tye,
{\em Classification of Closed Fermionic String Models}, Phys.\
Rev.\ {\bf D34} (1986) 3794.
\bibitem{rohm} J.\ Scherk and J.\ Schwarz, Nucl.\ Phys.\ {\bf B153}
(1977) 61. R.\ Rohm, {\em Spontaneous Supersymmetry Breaking in
Supersymmetric String Theories}, Nucl.\ Phys.\ {\bf B237} 553
(1984). C.\ Kounnas and M.\ Porratti, Nucl.\ Phys.\ {\bf B310}
(1988) 355.
\bibitem{dudas} I.\ Antoniadis, E.\ Dudas, and A.\ Sagnotti,
{\em Supersymmetry Breaking, Open Strings, and M Theory},
hep-th/9807011. K. Dienes, E.\ Dudas, T.\ Gherghetta, and A.\
Riotto, {\em Cosmological Phase Transitions, and Radius
Stabilization in Higher Dimensions}, hep-th/9809406.
\bibitem{update}E.\ Dudas, M.\ Nicolasi, G.\ Pradisi, and A.\
Sagnotti, {\em On Tadpoles and Vacuum Redefinitions in String
Theory}, hep-th/0410101.
\bibitem{gv} P. Ginsparg and C. Vafa,
{\em Toroidal Compactifications of Nonsupersymmetric
Heterotic Strings}, Nucl.\ Phys.\ {\bf B289} (1987) 414.
\bibitem{itoyama}
H. Itoyama and T. Taylor, {\em Supersymmetry Restoration in the
Compactified $O(16)$$\times$$O(16)$ Heterotic String Theory},
Phys.\ Lett.\ {\bf B186} (1987) 129.
\bibitem{dienes}J.\ Blum and K.\ Dienes, {\em Duality without
Supersymmetry: The Case of the $SO(16)$$\times$$SO(16)$ String},
Phys.\ Lett.\ {\bf B414} (1997) 260, hep-th/9707148.
\bibitem{bergman}O.\ Bergman and M.\ Gaberdiel, {\em Dualities of Type 0 Strings}, JHEP 9907 (1999) 022,
hep-th/9906055.
\bibitem{abel}A. Bytsenko, L. Granda and S. D. Odintsov,
Mod Phys Lett A11 (1996) 2525. S.\ Abel, J.\ Barbon, I.\ Kogan,
and E.\ Rabinovici, {\em String Thermodynamics in D-Brane
Backgrounds}, JHEP 9904 (1999) 015. R.\ Easther, B.\ Greene, M.\
Jackson, and D.\ Kabat, {\em Brane Gases in the Early Universe:
Thermodynamics and Cosmology}, JCAP 0401 (2004) 006.
\bibitem{witsei}A.\ Abouesaood, C.\ Nappi, and S.\ Yost, {\em Open String Theory
in Background Gauge Fields}, Nucl.\ Phys.\ {\bf B280} (1987) 599.
N.\ Seiberg and E.\ Witten, {\em String Theory and Noncommutative
Geometry}, hep-th/9908142. S.\ Carroll, J.\ Harvey, V. Kostelecky,
C.\ Lane, and T.\ Okamoto, {\em Noncommutative Field Theory and
Lorentz Violation}, Phys.\ Rev.\ Lett.\ {\bf 87} 141601,
hep-th/0105082.
\bibitem{malda}J.\ Maldacena, {\em The Large N Limit of
Superconformal Field Theories and Supergravity}, Adv.\ Theor.\
Math.\ Phys.\ {\bf 2} (1998) 505, hep-th/9711200.
\bibitem{witads}E.\ Witten, {\em Anti-de Sitter
Space, Thermal Phase Transitions, and Confinement in Gauge
Theories}, Adv.\ Theor.\ Math.\ Phys. {\bf 2} (1998) 505-532,
hep-th/9803131. Sec 2 reviews the basics of finite temperature
supersymmetric gauge theory phenomena.
\bibitem{sugimoto}S.\ Sugimoto, {\em Anomaly Cancellations in the
Type I D9-antiD9 System and the USp(32) String Theory}, Prog.\
Theor.\ Phys.\ {\bf 102} (1999) 685, hep-th/9905159.
\bibitem{schw}J.\ Schwarz and E.\ Witten, {\em Anomaly Analysis of
Brane Antibrane Systems}, JHEP 0103 (2001) 032, hep-th/0103099.
\bibitem{sundborg}See the recent works: B.\ Sundborg,
{\em The Hagedorn Transition, Deconfinement, and N=4 SYM}, Nucl.\
Phys.\ {\bf B573} (2000) 349, hep-th/9908001. O.\ Aharony, J.\
Marsano, S.\ Minwalla, K.\ Papadodimas, M.\ van Raamsdonk, {\em
The Hagedorn Transition in Weakly Coupled Large N Gauge Theories},
hep-th/0310285.
\bibitem{newtd}K.\ Dienes and M.\ Lenneck, hep-th/0312173, 0312216, 0312217,
see footnote [7].
\bibitem{garyp}G.\ Horowitz and J.\ Polchinski, {\em A
Correspondence Principle for Black Holes and Strings}, Phys.\
Rev.\ {\bf D55} (1997) 6189, hep-th/9612146.
\bibitem{rab}J.\ Barbon and E.\ Rabinovici, {\em Touring the Hagedorn
Ridge}, hep-th/0407236.
\bibitem{bernard}C.\ Bernard, {\em Feynman Rules for Gauge Theories at
Finite Temperature}, Phys.\ Rev.\ {\bf D9} 3312 (1974).
\bibitem{svet}A.\ M.\ Polyakov, {\em Thermal Properties of Gauge Fields and
Quark Liberation}, Phys.\ Lett.\ {\bf B72} (1978) 472. L.\
Susskind, Phys.\ Rev.\ {\bf D20} (1979) 2610. G.\ t'Hooft, Nucl.\
Phys.\ {\bf B138} (1978) 1. L.\ Mc Lerran and B.\ Svetitsky,
Phys.\ Rev.\ {\bf D24} (1981) 450. D. Gross, L. Yaffe, and M.
Perry, {\em Instability of Flat Space at Finite Temperature},
Phys.\ Rev.\ {\bf D25} (1982) 330.
\bibitem{karsch}See the recent works: O. Kaczmarek, F.
\ Karsch, P.\ Petreczky, and F.\ Zantow, {\em Heavy
Quark-Antiquark Free Energy and the Renormalized Polyakov Loop},
Phys.\ Lett.\ {\bf B543} (2002) 41, hep-lat/0312015. M.\ Quandt,
H.\ Reinhardt, and M.\ Engelhart, {\em Center Vortex Model for the
Infrared Sector of SU(3) YM Theory: Vortex Free Energy},
hep-lat/0412033, and citations thereof.
\bibitem{green}M.\ B.\ Green, {\em
Wilson-Polyakov Loops for Critical Strings and Superstrings at
Finite Temperature}, Nucl.\ Phys.\ {\bf B381} (1992) 201.
\bibitem{giov}M.\ Gasperini and G.\ Veneziano, {\em Pre-Big Bang Scenario
in String Cosmology}, hep-th/0207130. hep-th/9802057. F.\ Quevedo,
{\em Lectures on String/Brane Cosmology}, hep-th/0210292.
\bibitem{cosmic}E.\ Witten, Phys.\ Lett.\ {\bf B153} (1985) 243.
A.\ Vilenkin and E.\ P.\ S.\ Shellard, {\em Cosmic Strings and
Other Topological Defects}, Cambridge (1994). N.\ Jones, H.
Stoica, and S.-H.\ Henry Tye, JHEP 07 (2002) 051, hep-th/0203163.
S.\ Sarangi and S.\-H.\ Henry Tye, Phys.\ Lett.\ {\bf 536} (2002)
185, hep-th/0204074. E.\ Copeland, R.\ Myers, and J.\ Polchinski,
JHEP 06 (2004) 013, hep-th/0312067. J.\ Polchinski, {\em Cosmic
Superstrings Revisited}, hep-th/0410082.
\bibitem{dbrane}J.\ Polchinski, {\em Dbranes and Ramond-Ramond
Charge}, Phys.\ Rev.\ Lett.\ {\bf 75} (1995) 2724.
\bibitem{polwit}
J.\ Polchinski and E.\ Witten, {\em Evidence
for Heterotic-Type I String Duality}, Nucl.\ Phys.\ {\bf B460}
(1995) 525.
\bibitem{dkps}M. Douglas, D.\ Kabat, P.\ Pouliot, and S.\ Shenker,
{\em Dbranes and Short Distances in String Theory}, Nucl.\ Phys.\
{\bf B485} (1997) 85.
\bibitem{cmnp}A.\ Cohen, G.\ Moore, P.\ Nelson, and
J.\ Polchinski, {\em An Off-shell Propagator for String Theory},
Nucl.\ Phys.\ {\bf B267} (1986).
\bibitem{pair}S.\ Chaudhuri, Y.\ Chen, and E.\ Novak,
{\em Pair Correlation Function of Wilson Loops}, Phys.\ Rev.\ {\bf
D62} (2000) 026004.
\bibitem{pairf} S.\ Chaudhuri and E.\ Novak,
{\em Supersymmetric Pair Correlation Function of Wilson Loops},
Phys.\ Rev.\ {\bf D62} (2000) 046002. \bibitem{lower3} S.\
Chaudhuri, {\em A Proposal for Altering the Unification Scale in
String Theory}, Phys.\ Lett.\ {\bf B546} (2002) 108,
hep-th/0203058.
\bibitem{dhoker}E.\ D'Hoker and D.\ H.\ Phong, {\em Two Loop
Superstrings: Main Formulas}, Phys.\ Lett.\ {\bf B529} (2002) 241,
hep-th/0110247, and citations thereof. K.\ Aoki, E.\ D'Hoker, and
D.\ H.\ Phong, {\em Two Loop Superstrings on Orbifold
Compactifications}, hep-th/0312181.
\bibitem{sussk}L.\ Susskind, {\em Some Speculations about Black
Hole Entropy in String Theory}, hep-th/9309145; {\em The World as
a Hologram}, hep-th/9409089. G\ 't Hooft, {\em Dimensional
Reduction in Quantum Gravity}, gr-qc/9310006. R.\ Bousso, {\em The
Holographic Principle}, Rev.\ Mod.\ Phys.\ {\bf 74} (2002) 825,
hep-th/0203101.
\bibitem{mtheory}S.\ Chaudhuri, {\em Spacetime Reduction of Large N Flavor Models:
A Fundamental Theory of Emergent Local Spacetime Geometry?},
hep-th/0408057.
\bibitem{hodge}S.\ Chaudhuri, {\em Electric-Magnetic Duality and
the Brane Spectrum of M Theory}, hep-th/0409033.
\bibitem{micron}S.\ Chaudhuri, {\em Microcanonical Ensemble of
Type I Strings}, hep-th/0502141.
\end{thebibliography}
\end{document}